\documentclass[11pt]{article}
\usepackage{makeidx}
\usepackage{setspace}
\usepackage{multirow}
\usepackage{multicol}
\usepackage{graphicx}
\usepackage{latexsym}
\usepackage{amsmath} 
\usepackage{amssymb}
\usepackage{ulem}
\usepackage{cite}

\title{One-dimensional quantum wires: 
A pedestrian approach to bosonization}
\newcommand{\styleapsboldfontOne}[1]{{\footnotesize \textbf{#1}}}

\newcommand{\be}{\begin{eqnarray}}
\newcommand{\ee}{\end{eqnarray}}
\newcommand{\simgt}{\,\hbox{\lower0.6ex\hbox{$\sim$}\llap{\raise0.6ex\hbox{$>$}}}\,}

\begin{document}
\begin{center}
{\LARGE One-dimensional quantum wires: \\
A pedestrian approach to bosonization\\ ~\\}
{\Large Sebastian Eggert\\} {
University~of Kaiserslautern\\ 
Dept.~of Physics, 
Erwin-Schr\"odinger Str.\\ D-67663 Kaiserslautern, Germany \\}
\end{center}
\bibliographystyle{apsrev}




\tableofcontents

\section{Introduction }

In these lecture notes we will consider systems in which the motion of
electrons is confined to one dimension (1D).  In these so-called
\textit{quantum wires }electron-electron interaction effects play an
important role because the restricted dimensions enhance the scattering
between the electrons and completely destroy the quasi-particle
picture.  New density wave excitations appear that are described by
bosonic operators.  Here we will develop this bosonic description,
following a \textquotedblleft{}pedestrian\textquotedblright{} approach
which does not require any previous knowledge in field theory methods. 
These notes therefore serve as a detailed introduction into
bosonization by carefully deriving the most fundamental formulas.  For
advanced topics we recommend to consult one of the more sophisticated
reviews on bosonization \cite{1,2,3,4,5}
for further reading.

\subsection{What are quantum wires?}

Before we address many-body effects, let us review some introductory
material in order to define what quantum wires are, what can be
measured in typical experimental setups and what theoretical models we
wish consider.

\subsubsection{Band structure }

\begin{figure}[ht]
\begin{center}
\includegraphics[width=128pt]{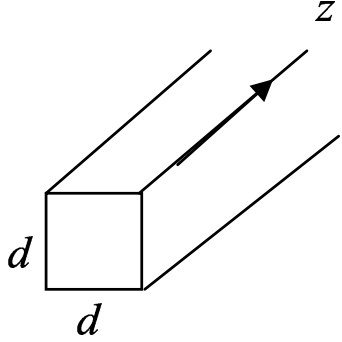}
\caption{Confinement in a square wire}
\label{f1}
\end{center}
\end{figure}
\begin{figure}[ht]
\begin{center}
\includegraphics[width=401pt]{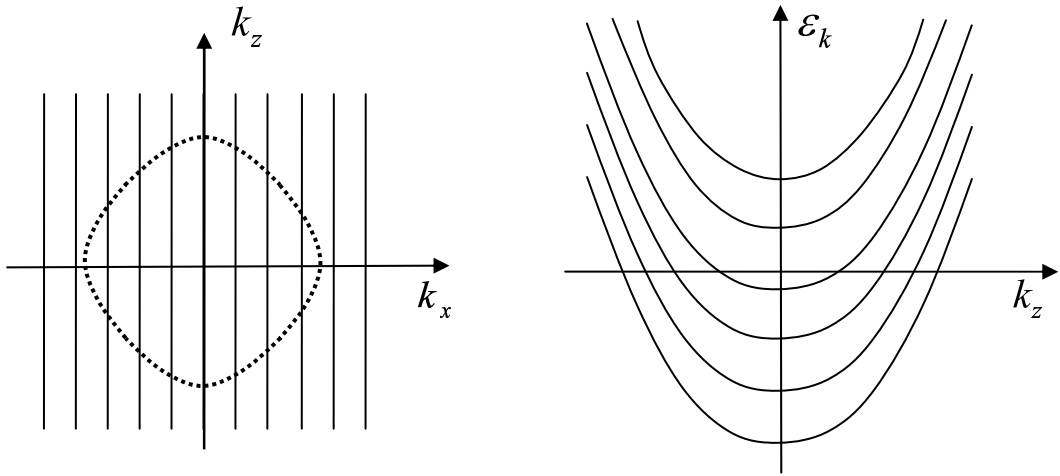}
\caption{Cross section of Fermi surface. Vertical lines are the 1D quantization 
lines.  Right:	Effective one-dimensional band-structure along the z-direction.  }
\label{f2}
\end{center}
\end{figure}

In the classical world we have some intuition on how to create a
one-dimensional transport channel for particles.  In order to make it
truly one-dimensional we could simply narrow a pipe until two particles
can no longer pass each other, for example by creating a narrow path
for pedestrians until they can only walk in single file.  It is easy to
imagine how a stop-and-go density wave will be created in such a
situation.  In quantum physics, particles are instead described by
wave-functions, which are known to be given by discrete standing waves
when we confine an electron with box-like boundary conditions as for
example shown in Fig. \ref{f1} for a simplified square wire.  The allowed
standing waves must have wave-numbers that fit in the square potential
along the $x-$ and $y-$directions, i.e.  
$d = n_x \frac{{\lambda _x }}{2} = n_x \frac{\pi }{k_x }$ 
and $d = n_y \frac{{\lambda _y }}{2} = n_y \frac{\pi }{{k_y }}$ 
while the motion in the z-direction is
unrestricted.  The allowed \begin{math} k-\end{math} vectors are
therefore along quantization lines which cut the band-structure of the
material of the wire.  This situation is depicted in Fig. \ref{f2}, where
only the $x-$ and $z-$direction is shown.  The original
Fermi surface of the material of the wire is schematically drawn dotted
in this plane.  Small energy excitations are now only possible along
the quantization lines and close to the Fermi surface.  Each
quantization line that crosses the Fermi surface therefore corresponds
to a \textquotedblleft{}channel\textquotedblright{}, which contributes
to the conductivity with quantized conductance as we will see later.

\begin{figure}
\begin{center}
\includegraphics[width=237pt]{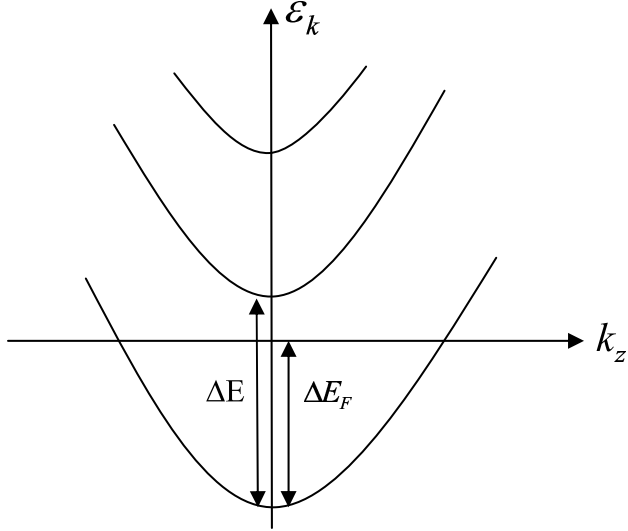}
\caption{Electron bands where the energy spacing is larger than the depth of Fermi sea.}
\label{f3}
\end{center}
\end{figure}

For a truly one-dimensional wire we would like to have only one single
channel that crosses the Fermi surface, i.e.~only two Fermi points
corresponding to left- and right-moving electrons.  For this to be
true, the energy spacing between the quantization lines must be larger
than the depth of the Fermi-surface as shown in Fig. \ref{f3}.  For an
order-of-magnitude estimate of the required length scale of
\begin{math} d\end{math} we can assume that the bandwidth is about a few
{eV} and the dispersion is given by an effective
mass \begin{math} E = \hbar^2 k^2 /2m^* \end{math}.  Together with
\begin{math} k = n\,\pi /d\end{math} the condition \begin{math}\Delta E
> \Delta E_F \end{math}  gives \begin{math}\Delta E = \frac{\hbar^2 }
{{2m^* }}\left( {\frac{\pi }
{d}} \right)^2  \simgt 1{\rm eV}\end{math} .  Using \begin{math}1Ry = \frac{{\hbar
^2 }}
{{2ma_0^2 }} \approx 13.6{\rm eV}\end{math}  and \begin{math}a_0  \approx
{\text{0}}{\text{.5{\AA}}}\end{math} , we finally arrive at the
estimate:
\begin{displaymath} d < \pi \sqrt {\frac{{\hbar ^2 }}
{{2m^* {\rm eV}}}}  = \pi \sqrt {\frac{{a_0^2 Ry}}
{{\rm eV}}}  = \pi \sqrt {13.6} \,a_0  \approx 0.5{\rm nm}\end{displaymath} 

We therefore find that diameters of one nanometer or less are required
in order to observe one-dimensional physics.  For semi-conductors the
bandwidth and the effective mass can be much smaller, so that a few
nanometers may be possible.  For metallic wires, however, essentially
single atomic chains are required.

\subsubsection{Creating quantum wires: Experiments }

The creation of one dimensional wires is therefore still a great
challenge even with today's semiconductor technology, which can create
structures that are only a few tens of nanometers in size.  However,
even if traditional processing techniques of lithography and etching
can be scaled down to about 10nm, the resulting structures are still
imprecise on a nanometer scale.  The resulting wire typically has
relatively uncontrolled wavy edges, which immediately leads to
localization in a one-dimensional system.  In order to produce a wire
that has perfect translational invariance, we therefore require
production mechanisms that give full control down to the atomic scale. 
We will briefly describe three of the most promising experimental
approaches below.

\begin{figure}
\begin{center}
\includegraphics[width=316pt]{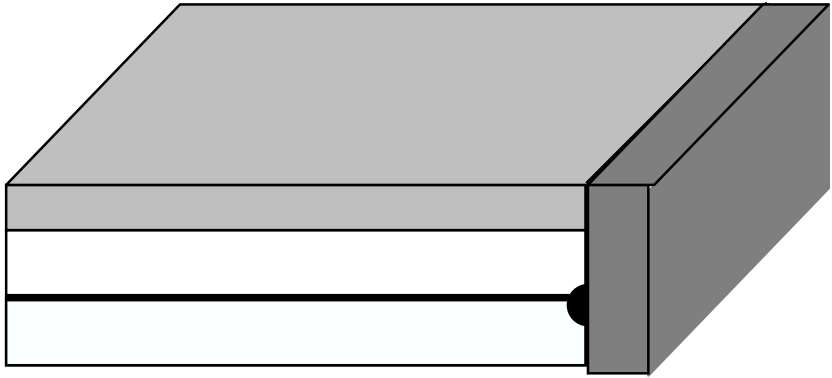}
\caption{1D wire created by cleaved edge overgrowth.\cite{6}}
\label{f4}
\end{center}
\end{figure}
Using molecular beam epitaxy it is possible to grow atomically perfect
structures layer by layer.  In this way it is now standard to create
high quality two-dimensional electron gas (2DEG) systems by use of an
inversion layer.  In the so-called cleaved edge overgrowth technique
such a sample is then cleaved in-situ\cite{6}.
The cleaved edge is typically along one of the crystal axis and
also atomically perfect.  If a gate is overgrown on the edge, it is
then possible to apply a suitable
potential that forces the mobile electrons to be trapped in the one
dimensional wire that is defined by the intersection of the 2DEG and
the cleaved edge as shown in Fig.~\ref{f4}.  A suppression of the density of
states which is consistent with the theory has been demonstrated in
such a setup\cite{7}.

\begin{figure}
\begin{center}
\includegraphics[width=316pt]{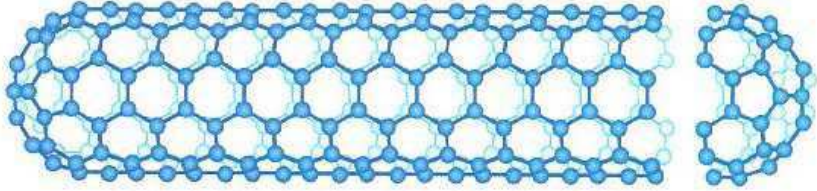}
\caption{A carbon nanotube.}
\label{f5}
\end{center}
\end{figure}
Another approach to create atomically perfect structures is to use
macromolecules that form a tubular structure.  The most famous examples
are the carbon nanotubes as shown in Fig.~\ref{f5}.  Even though the
detailed diameter and location of the tubes cannot be fully controlled
in the production process, the chemical composition guarantees that the
resulting tubes are atomically perfect structures on the nanoscale. 
Several experiments showed that single wall carbon nanotubes exhibit
the signatures of a one-dimensional many-body system \cite{8,9,10}
consistent with theory \cite{11,12,13}.

\begin{figure}
\begin{center}
\includegraphics[width=316pt]{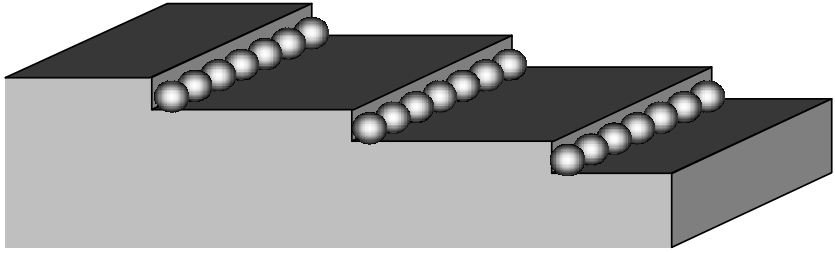}
\caption{Atomic Au chains deposited on Si step edges.\cite{14}}
\label{f6}
\end{center}
\end{figure}
Superstructures on extremely clean surfaces, like a stepped silicon
surface are a third example\cite{14}.
Cleaving a Si(111) surface at a slight angle creates a regular step
structure on which small amounts of gold can be deposited.  
The gold atoms tend to form
atomic chains along the steps of the silicon, a so-called Si(111) 5x1
Au structure as shown in Fig.~\ref{f6}.  Experiments were able to find the
signature of spin-charge separation using photoemission on such
samples\cite{14}.

Other possible experimental realizations of one-dimensional wires
include quantum hall edges, stretched point contacts, and intrinsically
quasi-low dimensional compounds.

\subsubsection{Models in second quantization}

\begin{figure}
\begin{center}
\includegraphics[width=216pt]{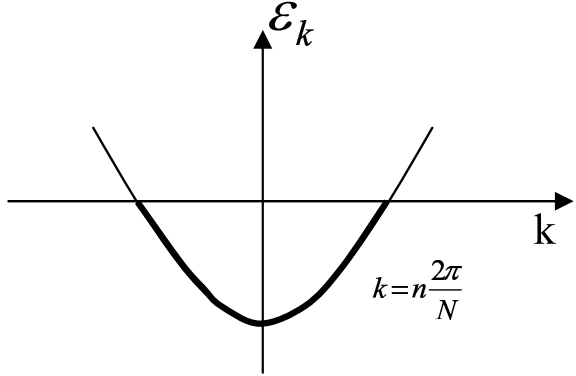}
\caption{Non-interacting electron dispersion.}
\label{f7}
\end{center}
\end{figure}

Let us now define the basic theoretical models for quantum wires.  The
Hamiltonian for an arbitrary band-structure of non-interacting fermions
as shown in Fig.~\ref{f7} can always be written in the formalism of second
quantization
\begin{equation}H = \sum\limits_{k,\sigma } {\varepsilon _k } c_{k,\sigma
}^\dagger  c_{k,\sigma } \label{H} \end{equation}
Here the operator \begin{math} c_{k,\sigma }^\dagger  \end{math} creates a
single electron \begin{math} \left| {\left. {k,\sigma } \right\rangle }
\right. = c_{k,\sigma }^\dagger  \left| {\left. 0 \right\rangle }
\right.\end{math}  in a Bloch eigenstate of the system and the
annihilation operator \begin{math} c_{k,\sigma } \end{math} annihilates
a particle.  These operators are defined for each spin index 
\begin{math} \sigma{}\end{math}=\begin{math}\uparrow{}\end{math},\begin{math}\downarrow{}\end{math}
and each band separately and are summed over independently in the
Hamiltonian.  Spin and band indices will be suppressed in what follows
until section 2.6.  For simplicity we can assume periodic Born-von
Karman boundary conditions over a finite length
$\ell=Na$, which leads to discrete
\begin{math} k\end{math} values of \begin{math}k = 2\pi n/N\end{math}. 
The usual anti-commutation relations for fermion operators must be
obeyed
\begin{eqnarray}
\left\{c_k^\dagger,  c_k^{\phantom{\dagger}}\right\} & = & c_k^\dagger  c_k^{\phantom{\dagger}}
+ c_k^{\phantom{\dagger}}  c_k^{\dagger} = \delta_{k,k'} \nonumber \\
\left\{c_k^\dagger,  c_k^\dagger\right\} & = & 0 = 
\left\{c_k^{\phantom{\dagger}}  c_k^{\phantom{\dagger}}\right\} 
\end{eqnarray}

This ensures that the wave-function that is created by these operators
is automatically anti-symmetric under particle exchange.  Also the
Pauli Exclusion Principle of no double occupancy \begin{math} c_k^\dagger 
c_k^\dagger   = 0\end{math}  is obeyed.  The number operator \begin{math}n_k 
= c_k^\dagger  c_k \end{math}  in the Hamiltonian  counts the number of
fermions in the state $|k\rangle$
with eigenvalues of 0 and 1, i.e.
\begin{equation}\left. {\left. {c_k^\dagger  c_k^{\phantom{\dagger}} } \right|0} \right\rangle  =
0\end{equation} and

\begin{equation}
 {{c_k^\dagger  c_k^{\phantom{\dagger}} } |k} \rangle  =  {c_k^\dagger  
c_k^{\phantom{\dagger}} c_k^\dagger  } | 0 \rangle  =  {\left( {c_k^\dagger  
\left\{ {c_k^{\phantom{\dagger}} ,c_k^\dagger  } \right\} - c_k^\dagger  c_k^\dagger  c_k } \right)}
| 0 \rangle  =  {c_k^\dagger  } | 0
\rangle  = | { k \rangle } 
\end{equation}

The Fourier transforms of the Bloch operators correspond to field
operators
\begin{equation}\psi (x_j ) = \frac{1}
{{\sqrt N }}\sum\limits_{k =  - \pi }^\pi  {e^{ikj} c_k }, \label{e3}
\end{equation}
which also obey canonical anti-commutation relations.  Here
\begin{math} x_j \end{math}\textit{=ja} with {$j=1,..,N$}
labels the locations in the one-dimensional lattice with lattice
spacing \begin{math} a\end{math}.  Therefore, \begin{math}\psi^\dagger  (x_j
)\end{math}  creates an electron in a localized Wannier state at lattice
site \begin{math} x_j \end{math}.  The field operator is periodic
\begin{math} \psi (x_j ) = \psi (x_{j + N} )\end{math} since
\begin{math} e^{ikN}  = 1\end{math} for \begin{math}k = 2\pi
n/N\end{math} .  The inverse Fourier transform is given by
\begin{equation}c_k  = \frac{1}
{{\sqrt N }}\sum\limits_{j =  1 }^N  {e^{-ikj} \psi(x_j) }.
\end{equation}

{~\\ \bf Example: Tight binding model}

In order to illustrate the use of second quantization in a simple
model, we consider hopping between neighboring orbitals along a chain. 
The field operator \begin{math} \psi (x_j )\end{math} now creates an
electron in the orbital wave-function \begin{math} \phi (\vec r - \vec
r_j )\end{math} , where \begin{math}\vec r_j \end{math} is the three
dimensional coordinate of the lattice site \begin{math} x_j \end{math}. 
The so-called tight binding Hamiltonian is represented by overlap
integrals t\begin{math}  =  - \int {d^3 } \vec r\phi (\vec r_j )\Delta
H\phi ^* (\vec r_{j + 1} )\end{math}  between neighboring orbitals,
which allows transitions of the electrons
(\textquotedblleft{}hopping\textquotedblright{}).  In second
quantization this Hamiltonian is given by
\begin{equation}H =  - t\sum\limits_j {\left( {\psi ^\dagger  (x_j )\psi (x_{j
+ 1} ) + \psi ^\dagger  (x_{j + 1} )\psi (x_j )} \right)}
\end{equation}
Using the Fourier transform (\ref{e3}), we get
\begin{equation}\begin{gathered}
H =  - \frac{t}
{N}\sum\limits_j {\sum\limits_k {\sum\limits_{k'} {\left( {e^{ - ikj}
c_k^\dagger  e^{ik'(j + 1)} c_{k'}  + e^{ - ik'(j + 1)} c_{k'}^\dagger  e^{ikj}
c_k } \right)} } }  \hfill \\
\quad  =  - t\sum\limits_k {\left( {e^{ik} c_k^\dagger  c_k  + e^{ - ik}
c_k^\dagger  c_k } \right)}  \hfill \\
\quad  =  - 2t\sum\limits_k {n_k \cos k}  \hfill \\
\end{gathered}
\end{equation}
where we have used the identity 
\begin{math} \sum\limits_{j = 1}^N {e^{ikj} }  = N\delta_{k} \end{math}
which follows from the
geometrical sum \begin{math} \sum\limits_{j = 0}^N {q^j  = \frac{{1 -
q^{N + 1} }}
{{1 - q}}} \end{math} .  This is an illustration of a band structure
\begin{math} \varepsilon _k  =  - 2t\cos k\end{math} with cosine
dependence where \begin{math} n_k  = c_k^\dagger  c_k^{\phantom{\dagger}} \end{math} counts
number of fermions in the state \begin{math} k\end{math}.

{~\\ \bf Interactions}

\begin{figure}
\begin{center}
\includegraphics[width=150pt]{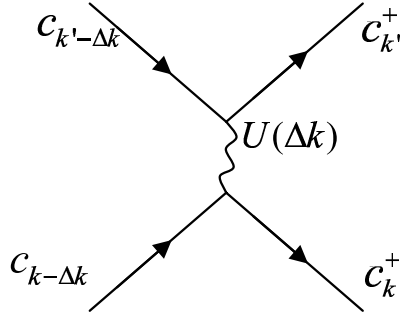}
\caption{The interaction corresponds to a scattering process.}
\label{f8}
\end{center}
\end{figure}
The understanding of electron-electron interaction effects is the main
motivation for bosonization.  The standard model of a generic
density-density interaction is represented by a two particle operator. 
If the interaction potential between electrons in Wannier states at
distance \begin{math} m\end{math} is given by
\begin{math} U(m)\end{math}, the corresponding interaction Hamiltonian
can be expressed in second quantization using the fermion density
operator \begin{math}\rho(x_j)= \psi ^\dagger  (x_j )\psi (x_j )\end{math} as
\begin{eqnarray}
H_{\rm int} & = &  \sum_{j=1}^N \sum_{m=1}^N \psi ^\dagger  (x_j )\psi (x_j )
U(m) \psi ^\dagger  (x_{j+m} )\psi (x_{j+m} ) \nonumber \\
&  =&  \frac{1}{{N^2 }}
\sum\limits_{j = 1}^N {\sum\limits_{m = 1}^N
{\sum\limits_{k,k'} {c_k^\dagger  } } } c_{k'}^{\phantom\dagger} e^{ - ikj} e^{ik'j}
U(m)\sum\limits_{k'',k'''} {c_{k''}^\dagger  } c_{k'''}^{\phantom\dagger} e^{ - ik''(j + m)}
e{}^{ik'''(j + m)} \label{e5}\end{eqnarray} 
The sum over \begin{math} j \end{math} results in a delta function with
the condition
${k-k'=k'''-k''= \Delta{}k}$,
so that,
\begin{eqnarray}H_{\operatorname{int} } &  = &  
\frac{1}
{N}\sum\limits_{m = 1}^N {\sum\limits_{k,k'',\Delta k} {c_k^\dagger  c_{k -
\Delta k}^{\phantom\dagger} } } U(m)c_{k''}^\dagger  c_{k'' + \Delta k}^{\phantom\dagger} e^{i\Delta km} \nonumber \\
& = & \frac{1}
{N} {\sum\limits_{k,k'',\Delta k} {c_k^\dagger  c_{k -
\Delta k}^{\phantom\dagger} } } U(\Delta k)c_{k''}^\dagger  c_{k'' + \Delta k}^{\phantom\dagger}  \label{e6}
\end{eqnarray}
where we have used the Fourier transform  and defined
\begin{math} U(\Delta k) = \sum\limits_m {e^{i\Delta km} U(m)}
\end{math} .  The two particle interaction therefore corresponds to a
scattering process as represented in the diagram in Fig.~\ref{f8}. 
Particles  \begin{math} c_{k' - \Delta k} \end{math} and 
\begin{math} c_{k - \Delta k} \end{math} are annihilated and after
exchanging  \begin{math} \Delta k\end{math} two new particles 
\begin{math} c_{k'}^\dagger  \end{math} and  \begin{math} c_k^\dagger  \end{math} are
created.

\subsection{What is measured?}

\subsubsection{Tunneling}

\begin{figure}
\begin{center}
\includegraphics[width=380pt]{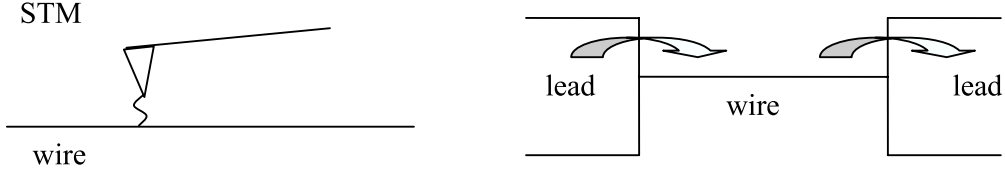}
\caption{Two tunneling examples: scanning tunneling microscopy and a conducting wire between two electrodes}
\label{f9}
\end{center}
\end{figure}
\begin{figure}
\begin{center}
\includegraphics[width=178pt]{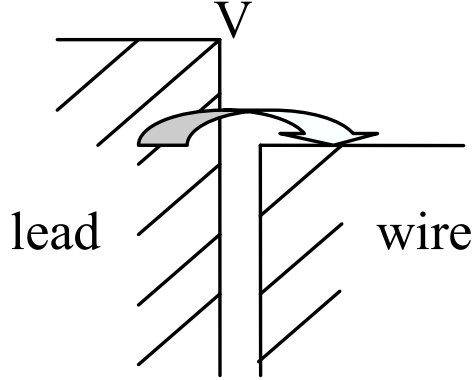}
\caption{Schematic diagram of tunneling from a lead/tip to a wire}
\label{f10}
\end{center}
\end{figure}

Tunneling is a common experimental setup for quantum wires.  One
example is scanning tunneling microscopy (STM), where tunneling occurs
across the gap between the wire and the STM tip.  A second example
would be a 1D wire, that is weakly connected to two leads at the ends,
where tunneling may occur.  In those experiments the current is
determined by the tunneling rate, which is given by Fermi's Golden
Rule
\begin{equation}\Gamma^+  (\omega ,x) = \frac{{2\pi t^2 }}
{Z}\sum\limits_{n,m} {e^{ - \beta E_m } \left| {\left\langle {n\left|
{\psi ^\dagger  (x)} \right|m} \right\rangle } \right|} ^2 \delta (\omega  -
E_n  + E_m ).\end{equation}
Here the probability of the transition from state  \begin{math} \left|
m \right\rangle \end{math}  to  \begin{math}\left| n \right\rangle
\end{math} by adding one particle is considered.  The tunneling
Hamiltonian between the lead/tip and the wire is assumed to
be \begin{math} H =  - t\left( {\psi ^\dagger  (x)\psi _{leads}  + h.c.}
\right)\end{math} and \begin{math} Z\end{math} is the partition function. 
The tunneling current can be calculated from the tunneling rates by
considering the empty and occupied states relative to the Fermi level
given by the Fermi-Dirac distribution, \begin{math} f(\omega ) = 1/(1 +
e^{\beta \omega } )\end{math} 
\begin{eqnarray}
I(V,x,\beta) = e\int_\infty^\infty d\omega \rho_{leads}(\omega-eV)
\left[ f(\omega-eV) \Gamma^+(\omega) \right.& -& \left.\left(1-f(\omega-eV)\right) 
\Gamma^-(\omega)\right] \nonumber \\
\rm (Tunneling\ in)&  - &\rm  (Tunneling\ out)  \label{I}
\end{eqnarray}
as illustrated in diagram \ref{f10}.  It is useful to express the tunneling
rate in terms of the so-called \textit{local spectral weight}, defined
as
\begin{equation} A(\omega ,x,\beta ) = \frac{{1 + e^{ - \beta \omega } }}
{Z}\sum\limits_{n,m} {e^{ - \beta E_m } } \left| {\left\langle
{n\left| {\psi ^\dagger  (x)} \right|m} \right\rangle } \right|^2 \delta
(\omega  - E_n  + E_m ) = \frac{{1 + e^{ - \beta
\omega } }}
{{2\pi t^2 }}\Gamma ^+, \label{A} \end{equation} 
assuming finite temperatures \begin{math} \beta  = 1/k_B T\end{math}. 
Inserting (\ref{A}) into the expression for the current (\ref{I}) and performing
straightforward calculations using the Fermi-Dirac distribution, we
arrive at a simple equation for \textit{I(V,x)} in terms of the local
spectral weight
\begin{equation} I(V,x,\beta ) = 2\pi t^2 e\int_{ - \infty }^\infty 
{d\omega \,\rho _{leads} (\omega  -  eV)} \left( {f(\omega  - eV) -
f(\omega )} \right)A(\omega ,x).\end{equation} 

Finally, in order to calculate the local spectral weight
\begin{math} A(\omega ,x)\end{math}, we can rewrite the delta function
as \begin{math} \delta (\omega ) = \frac{1}
{{2\pi }}\int_{ - \infty }^{ + \infty } {e^{\operatorname{i} \omega t}
dt}  = \operatorname{Re} \frac{1}
{\pi }\int_0^\infty  {e^{\operatorname{i} \omega t} } dt\end{math}, so
that from (\ref{A})
\begin{equation} 
A(\omega ,x) = \frac{1}
{Z}\sum\limits_{n,m} {(e^{ - \beta E_n }  + e^{ - \beta E_m } )}
\left| {\left\langle {n\left| {\psi ^\dagger  (x)} \right|m} \right\rangle }
\right|^2 \operatorname{Re} \frac{1}
{\pi }\int_0^\infty  {e^{i(\omega  - E_n  + E_m )t} dt}
\end{equation} 
Then, using \begin{math} e^{iHt} \left| m \right\rangle  = e^{iE_m t}
\left| m \right\rangle \end{math}, we arrive at the Green's function
representation
\begin{equation} \begin{gathered}
A(\omega ,x) = \operatorname{Re} \left( {\frac{1}
{{\pi Z}}\sum\limits_{n,m} {\int_0^\infty  {dte^{i\omega t} \left(
{e^{ - \beta E_m }  + e^{ - \beta En} } \right)} } \left\langle
{n\left| {\psi ^\dagger  } \right|m} \right\rangle \left\langle {m\left|
{e^{\operatorname{i} Ht} \psi e^{ - iHt} } \right|n} \right\rangle }
\right) \hfill \\
\quad \quad \quad \; = \operatorname{Re} \frac{1}
{\pi }\int_0^\infty  {dt\,e^{i\omega t} \left\langle {\psi (x,t)\psi ^\dagger
  (x,0) + \psi ^\dagger  (x,0)\psi (x,t)} \right\rangle }  \hfill \\
\quad \quad \quad \; = \frac{1}
{\pi }\operatorname{Im} \int_0^\infty  {G^R (t,x)e^{i\omega t} dt} 
\hfill \\
\end{gathered} \end{equation} 
where \begin{math}G^R (t,x)\end{math} is the
retarded Green's function
\begin{equation}
G^R(t,x) = -i\langle \left\{\psi(x,t),\psi^\dagger(x,0)\right\}\rangle \theta(t).
\label{G}
\end{equation}
Here the thermal expectation value \begin{math} \left\langle A
\right\rangle  = \sum\limits_n {e^{ - \beta E_n } } \left\langle n
\right|A\left| n \right\rangle /Z\end{math}  is used.  At zero
temperature the local spectral weight reduces to the\textit{ local
density of states} (LDOS)
\begin{equation}
\rho(\omega,x) = \sum_m\left|\langle m \left| \psi^\dagger (x) \right| 0\rangle \right|^2
\delta(\omega-\epsilon_m)=\frac{1}{\pi}{\rm Im} \int_0^\infty G^R(t,x)e^{i\omega t} dt
\label{rho}
\end{equation}
and the current is
\begin{equation} I(V,x,\beta ) = 2\pi t^2 e\int_0^{eV} {d\omega \,\rho
_{leads} (\omega  - eV)} \rho (\omega
,x)\end{equation} 

Since Green's functions such as (\ref{G})  play a central role in condensed
matter physics, it will be one of the main goals in the many-body
theory to calculate time correlation functions.

\subsubsection{Photoemission spectroscopy}

\begin{figure}
\begin{center}
\includegraphics[width=178pt]{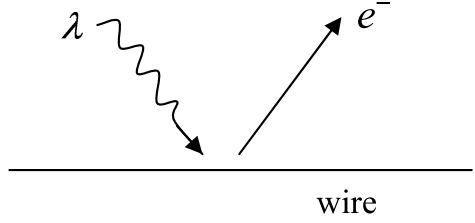}
\caption{Schematic diagram of photoemission}
\label{f11}
\end{center}
\end{figure}
Photoemission spectroscopy (PES) is a standard technique to determine
the electronic properties of condensed matter systems.  Photons with
definite energy and direction strike an object and electrons are
emitted via the photoelectric effect as shown in Fig.~\ref{f11}.  In
return, the intensity of the electrons as a function of absorbed energy
and momentum gives information about the object.  In inverse
photoemission electrons are used to probe the sample and the emitted
photons are analyzed correspondingly.  Interestingly the mathematical
description is very similar to the tunneling processes considered
above.  Just as before, we can argue that the emission rate of photons
\begin{math} \Gamma{}\end{math}$^{+}$ in inverse photoemission at a
given energy and wave-vector is related to the probability for adding a
corresponding electron into the system.  According to Fermi's Golden
Rule
\begin{equation} \Gamma ^+  (\omega ,q,\beta ) \propto \frac{{\text{1}}}
{{\text{Z}}}\sum\limits_{n,m} {e^{ - \beta E_m } } |\langle n|c_q^\dagger 
|m\rangle |^2 \delta (\omega  - E_n  + E_m ),\end{equation} 
where now the probability of the transition from state
\begin{math} \left| m \right\rangle \end{math} to \begin{math}\left| n
\right\rangle \end{math}  by adding one particle with wave-vector
\begin{math} q\end{math} is the relevant quantity.  In this case it is
useful to define the \textit{angle resolved spectral density}
\begin{equation} A(\omega ,q,\beta ) = \frac{{1 + e^{ - \beta \omega } }}
{Z}\sum\limits_{n,m} {e^{ - \beta E_m } } \left| {\left\langle
{n\left| {c_q^\dagger  } \right|m} \right\rangle } \right|^2 \delta (\omega 
- E_n  + E_m ) \propto \Gamma ^+  (\omega
,q).\end{equation} 

The angle integrated spectral density and the space integrated local 
spectral weight in equation (\ref{A}) are always the same and represent the total 
spectral weight as a function of $\omega$.
Following the analogous calculations as in the previous
section we can calculate the angle resolved spectral density in terms
of a retarded Green's function in momentum space
\begin{equation} A(\omega ,q,\beta ) =  - \frac{1}
{\pi }\operatorname{Im} \int_{\text{0}}^\infty  {dt{\kern 1pt}
\,{\text{e}}^{i\omega {\text{t}}} } G^R (t,q,\beta ),\end{equation} 
where
\begin{equation} G^R (t,q,\beta ) =  - i\left\langle {\left\{ {c_q (t),
c_q^\dagger  (0)} \right\}} \right\rangle.
\end{equation} 

\subsubsection{{Conductivity}}

\begin{figure}[ht]
\begin{center}
\includegraphics[width=178pt]{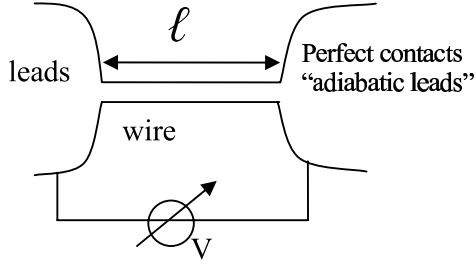}
\caption{Schematic diagram of metallic leads and a 1D wire}
\label{f12}
\end{center}
\end{figure}
When speaking about the physics of
\textquotedblleft{}wires\textquotedblright{}, 
it is natural to think about the corresponding conductivity.  However,
in order to apply a voltage and measure the current, perfect contacts
to external leads are required, which cannot be realized in most
experiments with truly one-dimensional wires.  Nonetheless, let us
imagine a setup in which electrons are adiabatically guided into a
single-channel wire of length \begin{math} \ell \end{math}\textit{
}without any scattering under an applied voltage \begin{math} V
\end{math} as shown in Fig.~\ref{f12}.

The standard theory of linear response gives the general conductivity
in terms of the Kubo formula
\begin{equation} \sigma (q,\omega ) = \frac{1}
{\omega }\int_0^\infty  {e^{i\omega t} } \left\langle {\left[
{J(q,t),J( - q,t)} \right]} \right\rangle .\end{equation} 

Naively the DC conductivity is then simply given by \begin{math} \sigma
(0,0)\end{math}, but since we are dealing with a finite wire it turns
out that it makes a difference if we first take the limit
\begin{math} \omega  \to 0\end{math} (current oscillating between the
leads) or the limit \begin{math} q \to 0\end{math} (current oscillating
within the wire).  In the latter case we obtain
\begin{equation} G(\omega ) = \mathop {\lim }\limits_{q \to 0} \sigma
(q,\omega ) = \frac{1}
{{\ell \omega }}\int_0^\ell  {dx} \int_0^\infty  {dte^{i\omega t} }
\left\langle {\left[ {J(x,t),J(0,0)} \right]} \right\rangle. \end{equation} 

\begin{figure}
\begin{center}
\includegraphics[width=178pt]{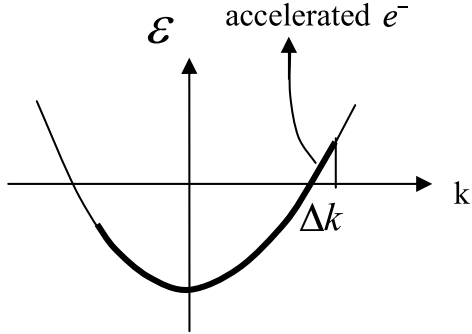}
\caption{Electrons are accelerated by the voltage into current carrying states.}
\label{f13}
\end{center}
\end{figure}

For interacting systems the outcome actually depends on the order of
limits.  However, we will not address this question here.  Instead we
consider a perfect ballistic non-interacting wire in order to show that
even in this simplest case the conductivity is finite and quantized. 
In fact, we can derive the conductivity without the calculation of
correlation functions by considering the semi-classical acceleration of
electrons in the field of an applied voltage \begin{math} E = V/\ell
\end{math}.   The crystal momentum changes with the applied force
\begin{math} \hbar \dot k =  - F = eE = eV/\ell \end{math}.  The current
is given by the number of electrons in current carrying states per unit
time.  The number of electrons that contribute to the current is simply
given by counting the states in an interval  \begin{math} \Delta
k\end{math} as  \begin{math}\# \;of\;e^ -   = \Delta k\,\ell /2\pi
\end{math} (see Fig.~\ref{f13}).  Since electrons are constantly ejected into
the right lead, the number of electrons in the current carrying states
remains finite and can be determined by the number of electrons that
are accelerated into those states per unit time via the following
simple calculation
\begin{displaymath} 
I = e\frac{{\# \,of\;e^ -  }}
{{\Delta t}} = e\frac{{\Delta k}}
{{\Delta t}}\frac{\ell }
{{2\pi }} = e\dot k\frac{\ell }
{{2\pi }} = \frac{{e^2 V}}
{{2\pi \hbar }} = GV
\end{displaymath} 
where \begin{math} G = e^2 /h\end{math} is the quantized conductance
per ballistic channel (\begin{math} 2e^2 /h\end{math} for electrons with
spin).  The corresponding energy is dissipated in the right lead where
the accelerated electrons quickly reach thermal equilibrium, while the
wire remains cold.  In general it is also possible to take a finite
transmission and several channels into account, which gives
\begin{displaymath} G = \sum\limits_{{\text{channels }}n} {t_n } \frac{{2e^2
}}
{h}\end{displaymath} 
where \begin{math} 0 < t_n  \leq 1\end{math} are the eigenvalues
of the transmission matrix \begin{math} T^\dagger  T\end{math} between the
\begin{math} n\end{math} different channels.  This is the famous
Landauer-B\"{u}ttiker formula for quantized conductance of wires and
point contacts.

\section{{Bosonic description}}

\subsection{{Linearization of the fermion dispersion}}

A bosonic description of the one-dimensional wire will be useful in
the treatment of interactions.  Let us first show in detail how a
non-interacting fermionic system given in equation (\ref{H})  can be expressed in
terms of a bosonic Hamiltonian in second quantization.  For this
equivalence to work we require the dispersion relation of the fermions
to be linear in the range of interest.  This is always approximately
true for excitations with small energy around the Fermi points. 
Therefore, we will restrict ourselves to consider low enough energies
so that we can describe the band structure in equation (\ref{H}) by an energy that
depends linearly on the wave vector \begin{math} k\end{math}, which is
the only approximation we will make.  The starting point is an
arbitrary free electron dispersion in equation (\ref{H})
\begin{equation} H = \sum\limits_{k,\sigma } {\varepsilon _k } c_{k,\sigma
}^\dagger  c_{k,\sigma }
\end{equation} 
At zero temperature all states between the Fermi wave-vectors
\textendash{}\textit{k$_{F}$} to \textit{k$_{F}$} are occupied.  The
number of particles in the Fermi sea is therefore given by
\begin{equation} n_0  = k_F N/\pi  - 1 \label{n0}\end{equation} 
if \textit{n$_{0}$} is even, or \textit{n$_{0}$=k$_{F}$}
\textit{N/\begin{math} \pi{}\end{math} }if \textit{n$_{0}$} is odd
(times two for electrons with spin).

\begin{figure}
\begin{center}
\includegraphics[width=214pt]{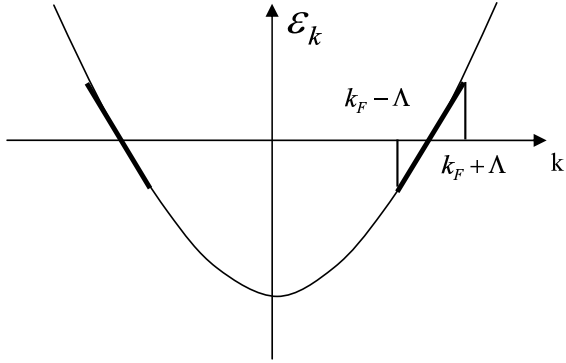}
\caption{Allowed k-values around $\pm k_F$} 
\label{f14}
\end{center}
\end{figure}
Since we are only interested in low energy excitations, we now
restrict the allowed wave-numbers to a range
\begin{math} \Lambda{}\end{math} around the Fermi points
$-k_F- \Lambda<k<-k_{F}+\Lambda{}$
and
$k_F- \Lambda<k<k_{F}+\Lambda{}$
as shown in Fig.~\ref{f14} corresponding to right- (\textit{k$_{F}$}) and
left-moving (\textendash{}\textit{k$_{F}$}) fermions, respectively.  The
range \begin{math} \Lambda{}\end{math} is determined by the region over
which the band structure is approximately linear and depends on the
model (typically corresponding to an energy range of about 1/10 of the
band width).  Keeping only states in this range the \textit{linearized
effective Hamiltonian} is given by
\begin{equation} H \approx \sum\limits_{k = k_F  - \Lambda }^{k_F  +
\Lambda } {\varepsilon _k c_k^\dagger  c_k }  + \sum\limits_{k =  - k_F  -
\Lambda }^{ - k_F  + \Lambda } {\varepsilon _k c_k^\dagger  c_k }
\end{equation} 
In this range the energy dispersion can be expanded around
\begin{math} \pm{}\end{math}\textit{k$_{F}$}
\begin{equation} \varepsilon _k  \approx \varepsilon _{k_F }  + \left.
{\left( {k - k_F } \right)\frac{{\partial \varepsilon _k }}
{{\partial k}}} \right|_{k_F }  + O\left[ {\left( {k - k_F } \right)^2
} \right]\ \ {\rm  for} \ \ k \approx k_F \end{equation}
for the \textquotedblleft{}right movers\textquotedblright{} and
analogously
\begin{equation} \varepsilon _k  \approx \varepsilon _{ - k_F }  + \left.
{\left( {k + k_F } \right)\frac{{\partial \varepsilon _k }}
{{\partial k}}} \right|_{ - k_F }  + O\left[ {\left( {k + k_F }
\right)^2 } \right]\ \ {\rm for}\ \  k \approx  - k_F
\end{equation} 
for the \textquotedblleft{}left movers\textquotedblright{}.  Here the
Fermi energy \begin{math} \varepsilon _{ - k_F }  = \varepsilon _{k_F } 
= 0\end{math}  must be zero since all occupied states have negative
energy and all empty states have positive energy at zero temperature. 
Defining a Fermi velocity
\begin{equation} \left. {v_F  = \left. {\frac{{\partial \varepsilon _k }}
{{\partial k}}} \right|_{k_F }  =  - \frac{{\partial \varepsilon _k }}
{{\partial k}}} \right|_{ - k_F }
\end{equation} 
we can therefore write \begin{math} \varepsilon _k  \approx v_F \left(
{k - k_F } \right)\end{math}  for right movers and
\begin{math} \varepsilon _k  \approx  - v_F \left( {k + k_F }
\right)\end{math}  for left movers.  It is useful to define new quantum
numbers \begin{math} \left| k \right| < \Lambda \end{math} shifted
relative to the Fermi points
\begin{math} \pm{}\end{math}\textit{k$_{F}$} and corresponding
\textit{left and right moving fermion operators}
\begin{eqnarray}
c_k^R & = &  c_{k_F+k} \nonumber \\
c_k^L & = &  c_{-k_F+k} \label{e13}
\end{eqnarray}
In this way the effective Hamiltonian  has the simple form
\begin{equation} H \approx \sum\limits_{k =  - \Lambda }^\Lambda  {v_F k}
\left( {c_k^{R \dagger } c_k^R  - c_k^{L \dagger} c_k^L }
\right)
\label{linH}
\end{equation} 

Accordingly, the original fermion field operator in equation  also
splits into a left and a right-moving part
\begin{equation} \begin{gathered}
\psi (x_j ) \approx \frac{1} {{\sqrt N }}\sum\limits_k {e^{ikj} c_k }  \hfill \\
\quad \quad \quad  = \frac{1}
{{\sqrt N }}\left( {\sum\limits_{k = k_F  - \Lambda }^{k_F  + \Lambda
} {e^{ikj} c_k }  + \sum\limits_{k = \Lambda  - k_F }^{ - k_F  -
\Lambda } {e^{ikj} c_k } } \right) \hfill \\
\quad \quad \quad  = \frac{1}
{{\sqrt N }}\sum\limits_k {\left( {e^{ik_F j} e^{ikj} c_k^R  + e^{ -
ik_F j} e^{  ikj} c_k^L } \right)}  \hfill \\
\quad \quad \quad  = \sqrt a \left( {e^{ik_F x_j /a} \psi _R (x_j )
+ e^{ - ik_F x_j /a} \psi _L (x_j )} \right) \hfill \\
\end{gathered}
\label{psi}
\end{equation} 
where we have defined left and right moving fermion fields on the
length of the wire \begin{math} \ell  = Na\end{math}
\begin{eqnarray}
\psi_R (x_j ) & = &  \frac{1} {{\sqrt \ell }}\sum\limits_{k=-\infty}^\infty 
{c_k^R e^{ikx_j/a} }  
\nonumber \\
\psi_L (x_j ) & = &  \frac{1} {{\sqrt \ell }}\sum\limits_{k=-\infty}^\infty 
{c_k^L e^{ikx_j/a} }  
\label{psiLR}
\end{eqnarray}
In those definitions it is useful to extend the range of summation to
infinity, because in this way an \textit{inverse} Fourier transform can
be defined as a continuous integral
\begin{eqnarray} 
c_k^{L/R}  = \frac{1}
{{\sqrt \ell  }}\int_0^\ell  {e^{ - ikx/a} \psi _{L/R} (x)}  dx \label{cLR}
\end{eqnarray}

One may object that we have now changed the range of summation to
infinity in equations (\ref{psi})  and (\ref{psiLR}), since the cutoff
\textit{\begin{math} \Lambda{}\end{math}} is supposed to be much smaller
than the bandwidth.  However, remember that
\textit{\begin{math} \Lambda{}\end{math}} also represents an upper limit
for the validity of the effective low energy description.  Indeed,
extending the states of the linear dispersion relation  to infinity or
leaving them out makes no difference if we always restrict ourselves to
low energies, since those states will then never take part in any
physical excitation.  Therefore, all states with
\textit{k\begin{math} >\end{math}\begin{math}\Lambda{}\end{math}} and
\textendash{}\textit{k\begin{math} <\end{math}\begin{math}\Lambda{}\end{math}}
are unphysical anyway.  By taking the summation range to infinity we
have effectively taken a continuous limit for the field operators,
which is mathematically more convenient.  However, this does
\textit{not} correspond to an additional approximation.  Note that the
anti-commutator
\begin{equation} \left\{ {\psi _R^\dagger  (x),\psi _R (y)} \right\} = \frac{1}
{\ell }\sum\limits_{kk'} {e^{ - ikx/a} e^{ik'y/a} } \left\{ {c_k^{R \dagger
} ,c_{k'}^R } \right\} = \frac{1}
{\ell }\sum\limits_{n =  - \infty }^\infty  {e^{ - i2\pi n(x - y)/\ell
} }  = \delta \left( {x - y} \right) \label{anticomm}\end{equation} 
is now normalized as a delta function.  In general, the normalization
would actually depend on the choice of the cut-off
\textit{\begin{math} \Lambda{}\end{math}}, but by including the
non-physical states a more convenient field operator with canonical
anti-commutation relations has been defined.

\subsection{{Excitation spectrum}}

In order to understand how the fermionic spectrum can be represented
by bosons we will consider individual excited states as e.g.~depicted
in Fig.~\ref{f15} on the right moving branch.
In terms of fermionic creation and annihilation operators the state is
expressed as (omitting the index \begin{math} R\end{math} for right
movers temporarily)
\begin{equation} c_0^{\phantom\dagger} c_1^\dagger  c_2^\dagger  c_4^\dagger  c_5^\dagger  c_8^\dagger  \left| {GS}
\right\rangle, \end{equation} 
where we use integer labeling relative to the highest occupied state
of the Fermi sea for simplicity\footnote{Note that the Fermi level can be chosen
to be in the middle between two states if $n_0$ is odd.  The highest
occupied state $n=0$ is at $k_F -\pi/N$
and $k$ is measured relative to $k_F$.}
\begin{equation} n = kN/2\pi  + \raise.5ex\hbox{$\scriptstyle
1$}\kern-.1em/
\kern-.15em\lower.25ex\hbox{$\scriptstyle 2$}. \label{e20} \end{equation}  
Note that the order of the operators matters, since they anti-commute.
Alternatively, we can construct this state in two parts:
\begin{figure}
\begin{center}
\includegraphics[width=379pt]{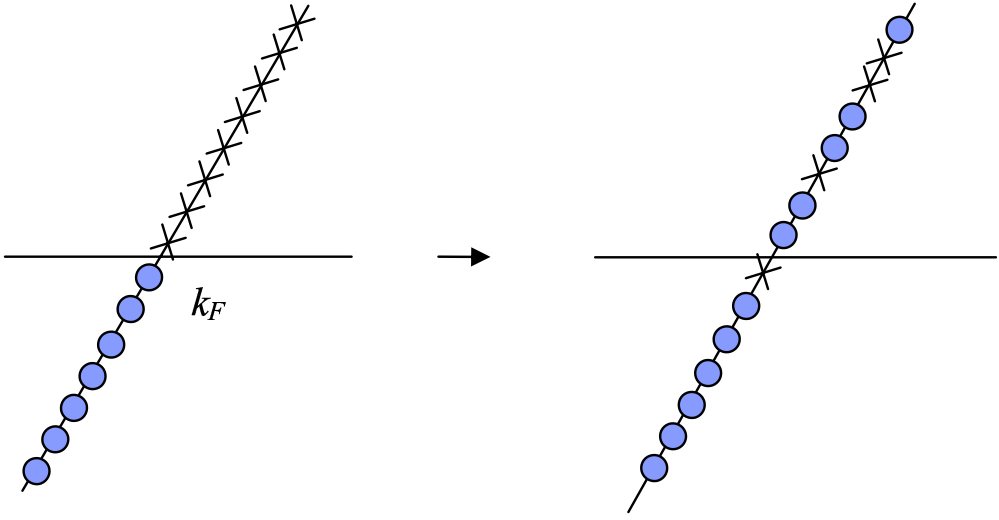}
\caption{Left: The ground state of the filled Fermi sea $|GS\rangle$.  
Right: Typical excited state.  The dots represent occupied states, 
crosses unoccupied states.}
\label{f15}
\end{center}
\end{figure}

\begin{figure}[ht]
\begin{center}
\includegraphics[width=180pt]{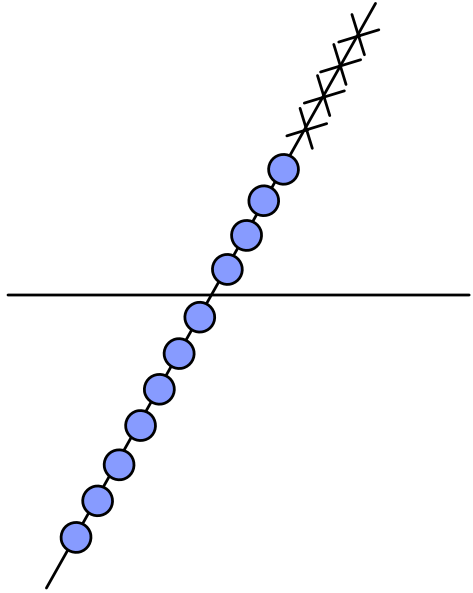}
\caption{Excited state after adding four fermions in the lowest unoccupied levels 
$c_1^\dagger c_2^\dagger c_3^\dagger c_4^\dagger |GS\rangle$ after step 1.}
\label{f16}
\end{center}
\end{figure}

1.\hspace{15pt}First we create four additional fermions in the lowest
unoccupied states \begin{math} c_1^\dagger  c_2^\dagger  c_3^\dagger  c_4^\dagger  \left|
{GS} \right\rangle \end{math}  as shown in Fig.~\ref{f16}.  The energy cost
for this is given by$^{1}$
\begin{displaymath} \frac{{2\pi v_F }}
{N}\left( {\frac{1}
{2} + \frac{3}
{2} + \frac{5}
{2} + \frac{7}
{2}} \right) = \frac{{2\pi v_F }}
{N} \cdot \frac{{16}}
{2},\end{displaymath} 
or in general by
\begin{equation} E_{n_R }  = \frac{{\pi v_F }}
{N}n_R^2, \label{e21}
\end{equation} 
where \textit{n$_{R}$} is the total number of additional right movers.
The energy cost is the same for removing \textendash{}\textit{n$_{R}$}
particles.

2.\hspace{15pt}Secondly we create particle-hole excitations by
shifting up the individual fermions (preserving their relative order). 
In our case we have to
\begin{itemize}
	\item shift the top fermion up by 4 steps with energy cost \begin{math} 1
\times 4 \times 2\pi v_F /N\end{math} ,
	\item shift zero fermions by 3 steps,
	\item shift the two next fermions by 2 steps each with energy \begin{math} 2
\times 2 \times 2\pi v_F /N\end{math} 
	\item finally shift the two next fermions up by 1 steps with energy
\begin{math} 2 \times 1 \times 2\pi v_F /N\end{math}
\end{itemize}
By this shifting procedure any particle-hole excited state can be
created.  If we now interpret that each shifting by
\begin{math} n\end{math} steps corresponds to a boson of level
\begin{math} n\end{math}, the particle-hole excitations in the part 2
above would correspond to
\begin{itemize}
	\item one excited boson on level 4, with energy \begin{math} 1 \times 4
\times 2\pi v_F /N\end{math} 
	\item no bosons on level 3
	\item two bosons on level 2 with energy \begin{math} 2 \times 2 \times 2\pi
v_F /N\end{math} 
	\item two bosons on level 1 with energy \begin{math} 2 \times 1 \times 2\pi
v_F /N\end{math} 
\end{itemize}
In this way any particle-hole excitation can in principle be
represented by bosonic occupation numbers.  This hand-waving picture is
almost the entire secret to bosonization.  The full story involves only
a slightly more complicated linear combination of shifted states as we
will see now.

\subsection{{Bosonic operators}}

From the previous section it is intuitively clear that
\textquotedblleft{}fermion shifting operators\textquotedblright{} may
be represented by bosons.  In this section we will make this
mathematically precise.  A good definition for a fermion shifting
operator is given by
\begin{equation} \rho _k^R  = \sum\limits_{k'} {c_{k' + k}^{R \dagger } c_{k'}^R
} \label{rhoR}\end{equation} 
This operator changes the wave-vector of right moving fermions by an
amount \begin{math} k\end{math}, summed over all possible states
\textit{k'}, resulting in a superposition of shifted states.  This is a
slight difference from the intuitive operations discussed in previous
section where we have only shifted the uppermost fermions at a time
(part 2), but clearly equation (\ref{rhoR})  amounts to the same basic idea.

The operator  is not hermitian.  In fact, the hermitian conjugate
changes the \begin{math} k\end{math}-vector by the opposite amount
\begin{equation} \left( {\rho _k^R } \right)^\dagger   = \sum\limits_{k' =  -
\infty }^{ + \infty } {c_{k'}^{R \dagger } c_{k' + k}^R  = \sum\limits_{k'}
{c_{k' - k}^{R \dagger} c_{k'}^R } }  = \rho _{ - k}^R
\end{equation} 
If \begin{math} \rho _k^R \end{math} with \begin{math}k>0\end{math}
creates an excitation as suggested in the previous section, the
hermitian conjugate annihilates it, which is indeed what we would
expect of bosonic operators.  In the last equation we have shifted the
summation variables, which is unproblematic because we have used the
convenient trick of extending the summation over an infinite range as
discussed above.

The final and most important step in order to relate the shifting
operators in (\ref{rhoR})  to bosons are the commutation relations.  Using the
definition (\ref{rhoR})  we get
\begin{equation} \begin{gathered}
\left[ {\rho _{ - k}^R ,\rho _{k'}^R } \right] =
\sum\limits_{k'',k'''}^{} {\left( {c_{k'' - k}^{R \dagger } c_{k''}^R c_{k'''
+ k'}^{R \dagger } c_{k'''}^R  - c_{k''' + k'}^{R \dagger } c_{k'''}^R c_{k'' -
k}^{R \dagger } c_{k''}^R } \right)}  \hfill \\
\quad \quad \quad \quad  = \sum\limits_{k'',k'''}^{} {\left( {c_{k''
- k}^{R \dagger } \left\{ {c_{k''}^R ,c_{k''' + k'}^{R \dagger } }
\right\}c_{k'''}^R  - c_{k''' + k'}^{R \dagger } \left\{ {c_{k'''}^R ,c_{k''
- k}^{R \dagger } } \right\}c_{k''}^R } \right.}  \hfill \\
\quad \quad \quad \quad \quad \quad \quad 
\left. 
-   {c_{k'' - k}^{R \dagger } 
c_{k''' + k'}^{R \dagger } 
c_{k''}^R 
c_{k'''}^R  + c_{k''' + k'}^{R \dagger }  
c_{k'' - k}^{R \dagger }  
c_{k'''}^R 
c_{k''}^R } \right)  \hfill \\
\quad \quad \quad \quad  = \sum\limits_{k''} {\left( {c_{k'' - k}^{R
\dagger } c_{k'' + k'}^R  - c_{k'' - k + k'}^{R \dagger } c_{k''}^R } \right)} 
\hfill \\
\quad \quad \quad \quad  = 0, \hfill \\
\end{gathered} \end{equation} 
where the two terms on the third line cancel and we 
have again shifted the summation variable in the last step
\begin{math} (k'' \to k'' - k)\end{math}.  Therefore we find that the
shifting operators generally commute.  However, if \begin{math} k =
k'\end{math}, the summation variable cannot simply be shifted because
in that special case the result corresponds to the subtraction of two
infinities
\begin{equation} \left[ {\rho _{ - k}^R ,\rho _k^R } \right] =
\sum\limits_{k''} {\left( {n_{k'' - k}  - n_{k''} } \right)}  = \infty  -
\infty
\end{equation} 
Indeed any arbitrary result can be produced by subtracting two
infinities!  However, we remember that these are not real infinities
(coming from non-physical states below the lower cutoff
\textit{k\begin{math} <\end{math}\textendash{}\begin{math}\Lambda{}\end{math}}
).  Therefore, using the fact that all states below the lower cutoff
always have to be occupied, 
we can cancel the unphysical infinities by using the cutoff at $-\Lambda$ as 
a reference point, since we require $n_k=1$ for $k<-\Lambda$
\begin{equation} \begin{gathered}
\left[ {\rho _{ - k}^R ,\rho _k^R } \right] = \sum\limits_{k'' <  -
\Lambda } {\left( {n_{k''-k} - n_{k''} } \right)}  + \sum\limits_{k'' \geq -
\Lambda } {\left( {n_{k'' - k}  - n_{k''} } \right)}  \hfill \\
\quad \quad \quad \quad  = \sum\limits_{k'' \geq - \Lambda  -
k} {n_{k''} }  - \sum\limits_{k'' \geq - \Lambda } {n_{k''} }  \hfill
\\
\quad \quad \quad \quad  = \sum\limits_{ - \Lambda  > k'' \geq
- \Lambda  - k} {n_{k''} }  \hfill \\
\quad \quad \quad \quad  = \frac{{kN}}
{{2\pi }} \hfill \\
\end{gathered}
\end{equation} 
where all terms in the first sum on the first line are zero 
and we have used the finite level spacing
\textit{k=\begin{math} 2\pi{}\end{math}n/N}.  In summary, we have
\begin{equation} \left[ {\rho _{ - k}^R ,\rho _{k'}^R } \right] = \delta
_{kk'} \frac{{kN}}
{{2\pi
}}
\label{boscomm}
\end{equation} 
which indeed corresponds to bosonic commutation relations.

For the bosonic operators to be useful we still must relate them to
the Hamiltonian.  Using equation (\ref{linH})  \begin{math} H = \sum\limits_{k'}^{}
{v_F k} '\left( {c_{k'}^{R \dagger } c_{k'}^R  - c_{k'}^{L \dagger} c_{k'}^L }
\right)\end{math}, the commutator with the shifting operator is given
by
\begin{eqnarray}
\left[H,\rho_k^R\right] 
& = & \sum_{k',k''} v_F k' \left[c_{k'}^{R \dagger}
c_{k'}^{R \phantom\dagger},c_{k''+k}^{R \dagger}c_{k''}^{R } \right] \nonumber \\
& = & \sum_{k',k''} v_F k' \left(
c_{k'}^{R \dagger} \left\{
c_{k'}^{R \phantom\dagger},c_{k''+k}^{R \dagger}\right\} c_{k''}^{R } 
-
c_{k''+k}^{R \dagger}
\left\{ 
c_{k''}^{R } 
,
c_{k'}^{R \dagger} 
\right\} 
c_{k'}^{R \phantom\dagger}
\right)
\nonumber \\
& = & \sum_{k'} v_F k' \left(
c_{k'}^{R \dagger}  c_{k'-k}^{R } 
-
c_{k'+k}^{R \dagger}
c_{k'}^{R \phantom\dagger}
\right)
\nonumber \\
& = & \sum_{k'} v_F  \left((k'+k)
c_{k'+k}^{R \dagger}  c_{k'}^{R } 
-
k' c_{k'+k}^{R \dagger}
c_{k'}^{R \phantom\dagger}
\right)
\nonumber \\
& = & v_Fk\rho_k
\label{Hrho}
\end{eqnarray}

It turns out that the two equations (\ref{boscomm})   and (\ref{Hrho})
are in fact sufficient to
determine the boson algebra and the bosonic representation of the
Hamiltonian.  In order to see this better we define conventional
creation boson operators for positive
\begin{math} k\end{math}-values\textit{ }(i.e. shifting particles up
\begin{math} k>0\end{math})
\begin{equation} b_k^{R \dagger }  = i\sqrt {\frac{{2\pi }}
{{kN}}} \rho _k^R, \label{bR}
\end{equation} 
while the hermitian conjugate annihilates an excitation by shifting
particles down
\begin{equation} b_k^R  =  - i\sqrt {\frac{{2\pi }}
{{kN}}} \rho _{ - k}^R
\end{equation} 
For left movers negative \begin{math} k\end{math}-values correspond to
an excitation, so that for \begin{math} k>0\end{math}
\begin{equation} \rho _{ - k}^L  = \sum\limits_{k'} {c_{k' - k}^{L \dagger}
c_{k'}^L }  = i\sqrt {\frac{{kN}}
{{2\pi }}} b_k^{L \dagger} \label{bL}
\end{equation} 
The phase in the definition is in fact arbitrary, but we choose
\textit{\begin{math} \pm{}\end{math}i} for later convenience.  Inserting
this definition in (\ref{boscomm}) and (\ref{Hrho})  we recognize the canonical commutation
relations
\begin{eqnarray} 
\left[ {b_k^R ,b_{k'}^{R \dagger } } \right] & = &  \delta _{kk'}, \label{e29} \\
\left[ {H ,b_{k}^{R \dagger } } \right] & = &  v_F k b_k^{R\dagger} \label{e30}
\end{eqnarray}    
and likewise for left-movers.  As shown in the treatment of the
quantum harmonic oscillator in any quantum mechanics book, operators
\textit{b$^{\dagger}$} and \begin{math} b\end{math} with such commutation
relations are called creation and annihilation operators and can be
used to build up the entire spectrum.  In our case, we have such an
oscillator spectrum for each \begin{math} k>0\end{math} separately,
describing the particle-hole excitations.  Now, also taking the energy
for adding/removing extra particles from equation (\ref{e21})  into account (see
Fig.~\ref{f16}), we arrive at the complete Hamiltonian in bosonic form
\begin{equation} H = v_F \sum\limits_{k > 0} {k\left( {b_k^{R \dagger } b_k^R  +
b_k^{L \dagger} b_k^L } \right)}  + \frac{{\pi v_F }}
{N}\left( {n_R^2  + n_L^2 }
\right)
\label{bosH}
\end{equation} 
where \begin{math} \left( {n_R^2  + n_L^2 } \right)\pi v_F /N\end{math}
is the energy in the first step of adding particles expressed in
counting operators \textit{n$_{R}$} and \textit{n$_{L}$} (so-called
\textit{zero modes}) and \begin{math} v_F k\left( {b_k^{R \dagger } b_k^R  +
b_k^{L \dagger} b_k^L } \right)\end{math}  corresponds to the second step of
shifting of particles (so-called \textit{oscillator modes}).

\subsection{{Fermion operators in terms of bosons}}

We have achieved the most important step of expressing the Hamilton
operator  entirely in terms of bosons.  However, if we wish to
calculate any physical expectation value we must also be able to
express more general fermion operators in terms of the bosons.  The
bosons are well defined in terms of the fermions by equations (\ref{bR}-\ref{bL}),
but we now seek the inverse transformation.

\subsubsection{{Left- and right-moving fermion densities}}

Using the concept of left-and right moving fermion field operators in
equation (\ref{psiLR}) we can immediately define a corresponding left and right
moving fermion density
\begin{equation} \rho _R^{\phantom\dagger} (x) = \psi _R^\dagger  (x)\psi _R
(x).\end{equation} 
Technically this operator is divergent because the ground state
expectation value is 
\begin{equation} \left\langle {\rho _R^{\phantom\dagger} (x)} \right\rangle  = \frac{1}
{\ell }\sum\limits_{k,k'} {e^{ - ikx/a} e^{ik'x/a} \left\langle
{c_k^{R \dagger } c_{k'}^R } \right\rangle }  = \frac{1}
{\ell }\sum\limits_{k =  - \infty }^0 {\left\langle {c_k^{R \dagger } c_k^R
} \right\rangle }  = \infty,
\end{equation} 
where we have used (\ref{psiLR}).
However, we understand again that this infinity comes from introducing
unphysical states below the lower cutoff
\textit{\begin{math}k <-\Lambda{}\end{math}}
as discussed above.  Since those states are always occupied, the same
divergence will appear whenever this operator is applied on any physical
state.  However, we can simply remove the divergence by subtracting the
ground state expectation value
\begin{equation} :\rho _R^{\phantom\dagger} (x):\; = \psi _R^\dagger  (x)\psi _R (x) -
\left\langle {\rho _R (x)} \right\rangle
\end{equation} 
This procedure of subtracting the ground state expectation value is
called \textit{normal ordering} and is indicated by the dots at the
sides.  This definition is indeed useful because we are actually only
interested in excitations on top of the well-defined ground state,
namely the filled Fermi-sea.  The normal ordered left- and right-moving
densities therefore measure density-fluctuations relative to the ground
state.  Instead of subtracting the ground state expectation value, we can 
equivalently always re-order 
the plane-wave fermionic operators \begin{math} c_k^{R \dagger
} \end{math}  and \begin{math}c_k^R \end{math}, so that the annihilation
operators are to the right for \begin{math} k>0\end{math} and to the
left for \begin{math} k\leq{}0\end{math} while keeping possible minus
signs from the anti-commutator, i.e.
\begin{eqnarray}
: c_k^{R\dagger}c_{k'}^R: = - : c_{k'}^{R}c_k^{R\dagger}: = \left\{
\begin{array}{ll}
c_k^{R\dagger}c_{k'}^R & {\rm for} \ k>0 \\
-c_{k'}^{R}c_k^{R\dagger} & {\rm for}\  k\leq 0\\
\end{array}
\right.
\label{order}
\end{eqnarray}
Note, that the normal ordering can be omitted if
\begin{math}k \not =k'\end{math}, since \begin{math}:c_k^{R \dagger }
c_{k'}^R :\; = \;c_k^{R \dagger } c_{k'}^R  =  - c_{k'}^R c_k^{R \dagger }
\end{math}  already follows from the regular anti-commutation relations 
in that case.  For normal ordered left movers the annihilation
operators are to the left for \begin{math} k>0 \end{math} and to the
right otherwise.

We now consider the Fourier transformation of the left- and
right-moving densities, first assuming {\begin{math}k \not
=0 \end{math}}.  By using (\ref{psiLR}) and \begin{math}\int_0^\ell 
{dx\,e^{ikx/a}  = \ell \delta _{k,0} } \end{math}  we find
\begin{equation} \begin{gathered}
\int_0^\ell e^{ikx/a} {dx:\rho _R (x):}  = \frac{1}
{\ell }\int_0^\ell  {dx\sum\limits_{k''k'} {e^{ikx/a} e^{ - ik''x/a} e^{ik'x/a}
:c_{k''}^{R \dagger } c_{k'}^R } } : \hfill \\
\quad \quad \quad \quad \ \quad \quad \quad \quad \,\;{\kern 1pt}  
= \sum\limits_{k'} {:c_{k+k'}^{R \dagger } c_{k'}^R } : 
= \sum\limits_{k'} {c_{k+k'}^{R \dagger } c_{k'}^R }  
\hfill \\
\quad \quad  \quad \quad \  \quad \quad \quad \quad \,\;{\kern 1pt}  = \rho_k^R  \hfill \\
\end{gathered}
\label{intrho}
\end{equation} 
On the second line we used the fact that the normal ordering in (\ref{order}) does
not affect the expression for \textit{\begin{math}k \not =0\end{math}}. 
We therefore find that the Fourier components of the right moving
densities correspond exactly to the fermion shifting operators defined
in (\ref{rhoR})!

For \begin{math} k=0\end{math} we have

\begin{equation} \begin{gathered}
\int_0^\ell  {dx:\rho _R (x):}  = \frac{1}
{\ell }\int_0^\ell  {dx\sum\limits_{k''k'} {e^{ - ik''x/a} e^{ik'x/a}
:c_{k''}^{R \dagger } c_{k'}^R } } : \hfill \\
\quad \quad \quad \quad \ \ \  \quad \,\;{\kern 1pt}  = \sum\limits_{k'}
{:c_{k'}^{R \dagger } c_{k'}^R } : \hfill \\
\quad \quad \quad \quad \ \ \quad \,\;{\kern 1pt}  = \;\sum\limits_{k' >
0} {c_{k'}^{R \dagger } c_{k'}^R }  - \sum\limits_{k' \leq 0} {c_{k'}^R
c_{k'}^{R \dagger } }  \hfill \\
\quad \quad \quad \quad \ \  \ \quad \,\;{\kern 1pt}  = n_R  \hfill \\
\end{gathered}
\label{intrho0}
\end{equation} 
We therefore find that the \textit{zero mode}
\textit{\begin{math} :\rho{}_{0}^{R}:\end{math} }corresponds to the
total number of right moving fermions relative to the ground state
\textit{\begin{math} :\rho{}_{0}^{R}:=n_{R}\end{math}} (the third
line corresponds to the number of excited particles
\textit{k'\begin{math} >0\end{math}} minus the number of holes
\textit{k'\begin{math} \leq{}0\end{math}}).  In terms of the bosons in 
(\ref{bR}) the equations (\ref{intrho})  and (\ref{intrho0})  can be summarized as
\begin{equation} \int_0^\ell  {dx\,e^{ikx/a} :\rho _R (x):}  =
\sum\limits_{k'} {:c_{k' + k}^{R \dagger } c_{k'}^R : = \left\{
\begin{gathered}
- i\sqrt {\frac{{kN}}
{{2\pi }}} b_k^{R \dagger } {\text{ for }}k > 0 \hfill \\
n_R \ \ \ \ \ \ \ \ \ \ \ \   {\text{                  for }}k = 0 \hfill \\
i\sqrt {\frac{{kN}}
{{2\pi }}} b_k^R \ \ \ \ {\text{    for }}k < 0{\text{ }} \hfill \\
\end{gathered}  \right.} {\text{
}}\end{equation} 
We now make use of the inverse Fourier transform of equations (\ref{intrho})  
and (\ref{intrho0}), which is given by (\begin{math} k = 2\pi n/N\end{math})
\begin{equation} \begin{gathered}
:\rho _R (x):\,\, = \frac{1}
{\ell }\sum\limits_{k} :\rho_k^R:e^{-ikx/a}
= \frac{1}
{\ell }\sum\limits_{k > 0} {\sqrt {\frac{{kN}}
{{2\pi }}} } \left( {ib_k^{R } e^{ikx/a}  - ib_k^{R \dagger} e^{ - ikx/a} }
\right) + \frac{{n_R }}
{\ell } \hfill \\
\quad \quad \;\quad \, = \frac{1}
{\ell }\sum\limits_{n = 1}^\infty  {\sqrt n } \left( {ib_n^{R }
e^{i\frac{{2\pi }}
{\ell }nx}  - ib_n^{R \dagger} e^{ - i\frac{{2\pi }}
{\ell }nx} } \right) + \frac{{n_R }}
{\ell } \hfill 
\end{gathered}
\label{modeR}
\end{equation} 
For the left moving density we can make the analogous calculations to
find
\begin{equation} \begin{gathered}
:\rho _L (x):\,\, = \frac{1}
{\ell }\sum\limits_{k > 0} {\sqrt {\frac{{kN}}
{{2\pi }}} } \left( {ib_k^{L \dagger} e^{ikx/a}  - ib_k^L e^{ - ikx/a} }
\right) + \frac{{n_L }}
{\ell } \hfill \\
\quad \quad \;\quad \, = \frac{1}
{\ell }\sum\limits_{n = 1}^\infty  {\sqrt n } \left( {ib_n^{L \dagger}
e^{i\frac{{2\pi }}
{\ell }nx}  - ib_n^L e^{ - i\frac{{2\pi }}
{\ell }nx} } \right) + \frac{{n_L }}
{\ell } \hfill \\
\end{gathered}
\label{modeL}
\end{equation} 
In summary, we have derived central bosonization formulas which can be
used in order to express fermionic densities in terms of bosons.

\subsubsection{{Boson field operator}}

It turns out to be useful to introduce a boson field operator
\textit{\begin{math} \phi{}\end{math}$_{R}$} in order to express the
bosonization formulas (\ref{modeR})  and (\ref{modeL}) more compactly and also in order to make
contact with conventional bosonic field theories.  Let us the define
the fields \textit{\begin{math} \phi{}\end{math}$_{R}$} and
\textit{\begin{math} \phi{}\end{math}$_{L}$}
\begin{equation} \begin{gathered}
\phi _R (x) = \phi _0^R  + Q_R \frac{x}
{\ell } + \sum\limits_{n = 1}^\infty  {\frac{1}
{{\sqrt {4\pi n} }}} \left( {e^{i\frac{{2\pi n}}
{\ell }x} b_n^R  + e^{ - i\frac{{2\pi n}}
{\ell }x} b_n^{R \dagger } } \right) \hfill \\
\phi _L (x) = \phi _0^L  + Q_L \frac{x}
{\ell } + \sum\limits_{n = 1}^\infty  {\frac{1}
{{\sqrt {4\pi n} }}} \left( {e^{ - i\frac{{2\pi n}}
{\ell }x} b_n^L  + e^{i\frac{{2\pi n}}
{\ell }x} b_n^{L \dagger} } \right) \hfill \\
\end{gathered} 
\label{phiLR}
\end{equation} 
in terms of the boson operators \begin{math} b_n^{R/L \dagger} \end{math}
and \begin{math} b_n^{R/L} \end{math} as defined in equations (\ref{bR}) and (\ref{bL}). 
The operators \begin{math} Q_R  = \sqrt \pi  n_R \end{math} and
\begin{math} Q_L  = \sqrt \pi  n_L \end{math} measure the right- and
left-moving particle number \textit{n$_{R}$} and \textit{n$_{L}$}.  The
operators \begin{math} \phi _0^R \end{math} and \begin{math}\phi
_0^L \end{math}  are defined as the canonical conjugate to the number
operators
\begin{eqnarray}
\left[\phi_0^R,Q_R\right] & = & -\frac{i}{2}
\nonumber \\
\left[\phi_0^L,Q_L\right] & = & \frac{i}{2}
\label{zerocomm}
\end{eqnarray}
The definition in equation (\ref{phiLR})  is useful, because we can immediately
express the fermion density in equations (\ref{modeR}) and (\ref{modeL})  as the derivative of the
boson field $\phi_{R}$ and $\phi_{L}$
\begin{equation} \begin{gathered}
:\rho _R (x):\; = \frac{1}
{{\sqrt \pi  }}\partial _x \phi _R (x) \hfill \\
:\rho _L (x):\; = \frac{1}
{{\sqrt \pi  }}\partial _x \phi _L (x) \hfill \\
\end{gathered}
\label{rhoLR}
\end{equation} 
Moreover, the Hamiltonian can also be expressed in terms of the boson
field as
\begin{equation} H = av_F \int_0^\ell  {dx\left( {(\partial _x \phi _R
)^2  + (\partial _x \phi _L )^2 } \right)} \label{bosH3}
\end{equation} 
Inserting the definition  and using \begin{math} \frac{1}
{\ell }\int_0^\ell  {dx\,e^{i\frac{{2\pi }}
{\ell }nx} }  = \delta _{n,0} \end{math}  we find that
\begin{equation} H = \frac{2 \pi v_F}{N} \sum\limits_{n=1}^\infty {n\left( {b_n^{R \dagger } b_n^R  +
b_n^{L \dagger} b_n^L } \right)}  + \frac{{v_F }}
{N}\left( {Q_R^2  + Q_L^2 }
\right) \label{bosH2}
\end{equation} 
which indeed is identical to the boson expression 
we found in equation
(\ref{bosH}).

\subsubsection{{Fermion field}}

In order to express arbitrary fermion operators and calculate Green's
functions, it is necessary to describe the fermion fields
\begin{math} \psi _R^{} (x)\end{math} and \begin{math}\psi _L^{}
(x)\end{math}  in terms of bosons.  We know that a right-moving fermion
field \begin{math} \psi _R^{} (x)\end{math} annihilates a fermion
locally at position \begin{math} x\end{math} and therefore changes the
density \begin{math} \rho _R (x) = \psi _R^\dagger  (x)\psi _R (x)\end{math}
at that point.  This action is reflected by the following commutation
relation
\begin{eqnarray}
\left[\psi_R(x),\rho_R(x')\right] 
& = & \psi_R(x)\psi_R^\dagger(x') \psi_R(x') - \psi_R^\dagger(x') \psi_R(x')\psi_R(x)
\nonumber \\
& =& \left\{ \psi_R(x),\psi_R^\dagger(x')\right\} \psi_R(x') \nonumber \\
& = & \delta(x-x') \psi_R(x)
\label{fermcomm}
\end{eqnarray}
Our goal is therefore to find a bosonic expression for
\begin{math} \psi _R^{} (x)\end{math} which reproduces the commutation
relation (\ref{fermcomm}).  The bosonic expression for the density \begin{math} :\rho _R
(x):\end{math}  is already known from equation (\ref{modeR}).

Before we go further, let us review a general relation for commutators
of two operators \begin{math} A\end{math} and \begin{math}B\end{math}. 
Assuming that \begin{math} \left[ {\left[ {A,B} \right],A} \right] =
0\end{math}  we can show that
\begin{eqnarray}
\left[e^A,B\right] = \sum_{n=1}^\infty \frac{1}{n!} \left[A^n,B\right]
=  \sum_{n=1}^\infty \frac{1}{n!} n A^{n-1}  \left[A,B\right] = e^A \left[A,B\right]
\label{expcomm}
\end{eqnarray}
Applying this equation to typical boson operators, i.e. \textit{A$=$b,
B$=$b$^{+}$} with $[A,B]=1$, we get
\begin{equation} \begin{gathered}
\left[ {e^{\alpha b} ,b^\dagger  } \right] = \alpha e^{\alpha b}  \hfill
\\
\left[ {e^{\beta b^\dagger  } ,b} \right] =  - \beta e^{\beta b^\dagger  } 
\hfill \\
\end{gathered}
\end{equation} 
The commutator with an exponential of boson operators is again
proportional to the exponential which is approximately what we want in
equation (\ref{fermcomm}).  We are therefore motivated to try the following ansatz to
describe the fermion field with an exponential of a linear combination
of boson operators with arbitrary coefficients
\textit{\begin{math} \alpha{}\end{math}$_{n}$} and
\textit{\begin{math} \beta{}\end{math}$_{n}$}
\begin{equation} \psi _R (x) \propto \exp \left( {\sum\limits_{n > 0}
{\left( {\alpha _n (x)b_n^R  + \beta _n (x)b_n^{R \dagger } } \right)} }
\right)
\label{ansatz}\end{equation} 
Using \begin{math} \left[ {b_n^R ,b_{n'}^{R \dagger } } \right] = \delta
_{n,n'} \end{math}  and  for each index \begin{math}n, n\end{math}', we
get
\begin{eqnarray}
\left[ \psi _R (x), b_n^{R\dagger} \right] & = & \alpha_n \psi_R(x) 
\nonumber \\
\left[ \psi _R (x), b_n^{R} \right] & = & -\beta_n \psi_R(x) 
\end{eqnarray}

Together with equation (\ref{modeR})  we can therefore calculate the commutator
\begin{equation} 
\begin{gathered}
\left[ {\psi _R (x),\rho _R (x')} \right] = \frac{1}
{\ell }\sum\limits_{n = 1}^\infty  {i\sqrt n \left( {e^{i\frac{{2\pi
n}}
{\ell }x'} \left[ {\psi _R (x),b_n^R } \right] - e^{ - i\frac{{2\pi
n}}
{\ell }x'} \left[ {\psi _R (x),b_n^{R \dagger } } \right]} \right)}  \hfill
\\
\quad \quad \quad \quad \quad \quad  =  - \frac{1}
{\ell }\sum\limits_{n = 1}^\infty  {i\sqrt n \left( {\beta _n
e^{i\frac{{2\pi n}}
{\ell }x'}  + \alpha _n e^{ - i\frac{{2\pi n}}
{\ell }x'} } \right)}  \psi _R (x) \hfill \\
\end{gathered}
\end{equation} 
So \begin{math} \left[ {\psi _R (x),\rho _R (x)} \right]\end{math} is
indeed proportional to \begin{math} \psi _R (x)\end{math} as desired. 
According to (\ref{fermcomm}) we must choose the coefficients so that we get a delta
function
\begin{equation}  - \frac{i}
{\ell }\sum\limits_{n = 1}^\infty  {\sqrt n \left( {\beta_n
e^{i\frac{{2\pi n}}
{\ell }x'}  + \alpha_n e^{ - i\frac{{2\pi n}}
{\ell }x'} } \right)}  = \delta (x - x').\end{equation} 
Comparing with the equation \begin{math} \frac{1}
{\ell }\sum\limits_{n =  - \infty }^{ + \infty } {e^{ - i\frac{{2\pi
}}
{\ell }n(x - x')} }  = \delta (x - x')\end{math}, we conclude that
\begin{equation} \alpha _n  = \frac{i}
{{\sqrt n }}e^{i\frac{{2\pi }}
{\ell }nx} \ \ \ \ \ \beta _n  = \frac{i}
{{\sqrt n }}e^{ - i\frac{{2\pi }}
{\ell }nx} \end{equation} 
This reproduces the delta function almost perfectly, up to the
\begin{math} n=0\end{math} term.  We therefore guess that we need the
zero mode in the exponential ansatz (\ref{ansatz}) as well.  From equations (\ref{zerocomm}) 
and (\ref{expcomm}) we know
\begin{equation} \left[ {e^{i\sqrt {4\pi } \phi _0^R } ,\frac{{Q_R }}
{{\sqrt \pi  }}} \right] = e^{i\sqrt {4\pi } \phi _0^R }
\label{e49}
\end{equation} 
Therefore,
\begin{equation} \psi _R (x) \propto \exp \left( {i\sqrt {4\pi } \phi
_0^R  + \sum\limits_{n > 0}^\infty  {\frac{i}
{{\sqrt n }}(e^{ - i\frac{{2\pi }}
{\ell }nx} b_n^{R \dagger }  + e^{i\frac{{2\pi }}
{\ell }nx} b_n^R )} } \right) \label{psiR1} \end{equation} 
fulfills the relation
\begin{equation} \begin{gathered}
\left[ {\psi _R (x),\rho _R (x)} \right] = \left[ {\psi _R
(x),\frac{{Q_R }}
{{\ell \sqrt \pi  }}} \right]\; + \,\;\frac{1}
{\ell }\sum\limits_{n = 1}^\infty  {i\sqrt n \left( {e^{i\frac{{2\pi
n}}
{\ell }x'} \left[ {\psi _R (x),b_n^R } \right] - e^{ - i\frac{{2\pi
n}}
{\ell }x'} \left[ {\psi _R (x),b_n^{R \dagger } } \right]} \right)}  \hfill
\\
\quad \quad \quad \quad \quad \quad  = \left( {\frac{1}
{\ell } + \frac{1}
{\ell }\sum\limits_{n = 1}^\infty  {(e^{i\frac{{2\pi n}}
{\ell }(x' - x)}  + e^{ - i\frac{{2\pi n}}
{\ell }(x' - x)} )} } \right)\  \psi _R (x) \hfill \\
\quad \quad \quad \quad \quad \quad  = \delta (x - x')\;\psi _R (x)
\hfill \\
\end{gathered} \end{equation} 
as required in equation (\ref{fermcomm}).  Including the
operator \textit{Q$_{R}$} in the exponential (\ref{psiR1}) does not change this
relation, so that we can arrive at a compact form by comparison with (\ref{phiLR})
\begin{equation}
\psi_R(x) \propto \exp\left(i \sqrt{4 \pi} \phi_R(x) \right)
\label{bosR}\end{equation}
which is the famous bosonization formula for fermionic fields. 
Following the analog calculations for left-movers we find that
\begin{equation} \psi _L (x) \propto \exp \left( { - i\sqrt {4\pi } \phi
_L (x)}
\right).\label{bosL}\end{equation} 

These two formulas will be useful when calculating Green's functions
and general correlation functions as we will see later.

\subsubsection{{Field commutators }}

So far we have not specified the overall proportionality constant of
the bosonization formulas (\ref{bosR}) and (\ref{bosL}).  
This can be fixed by using the proper
normalization introduced in equation (\ref{anticomm}), but it must be noted here that
even that normalization can in principle be chosen to be cutoff
dependent, so that we could equally well leave the overall constant
arbitrary, which is often done in the literature.  Nonetheless,
calculating the anti-commutator in bosonized form and calling the
proportionality constant \begin{math} C\end{math}, we get
\begin{equation} \begin{gathered}
\left\{ {\psi_R^\dagger  (x),\psi_R (y)} \right\} = C^2 \left\{ 
{e^{-i\sqrt {4\pi}\phi _R (x)} ,e^{i\sqrt {4\pi }\phi_R
(y)} } \right\} \hfill \\
\quad \quad \quad \quad \quad \quad \,\, = C^2 \;e^{ - i\sqrt {4\pi
} \left( {\phi_R (x) - \phi_R (y)} \right)} \left(
{e^{{\text{2}}\pi [\phi _R (x),\phi _R (y)]}  + e^{ -
{\text{2}}\pi [\phi_R (x),\phi _R (y)]} } \right), \hfill \\
\end{gathered} \label{C} \end{equation} 
where we have used the so-called Baker Hausdorf formula
\begin{equation} e^A e^B  = e^{A + B} e^{\frac{1}
{2}\left[ {A,B} \right]}  = e^B e^A e^{\left[ {A,B} \right]}
\label{baker}
\end{equation} 
(assuming \begin{math} \left[ {\left[ {A,B} \right],B} \right] = \left[
{\left[ {A,B} \right],A} \right] = 0\end{math} ).  Hence we need to
calculate the bosonic commutator (using (\ref{phiLR}) and (\ref{zerocomm}))
\begin{equation} 
\begin{gathered}
\left[ {\phi_R \left( x \right),\phi_R \left( y \right)}
\right] = \left[ {Q_R ,\phi _0^R } \right]\left( {\frac{{x - y}}
{\ell }} \right) \hfill \\
\quad \quad \quad \quad \quad \quad \quad \quad  + \sum\limits_{n,n' = 1}^\infty  {\frac{1}
{{4\pi \sqrt {nn'} }}} \left( {e^{i\frac{{2\pi }}
{\ell }\left( {nx - n'y} \right)} \left[ {b_n^R ,b_{n'}^{R \dagger } }
\right] + e^{ - i\frac{{2\pi }}
{\ell }\left( {nx - n'y} \right)} \left[ {b_n^{R \dagger } ,b_{n'}^R }
\right]} \right) \hfill \\
\quad \quad \quad \quad \quad \quad \,\,{\kern 1pt}  = i\frac{{x -
y}}
{{2\ell }} + \sum\limits_{n = 1}^\infty  {\frac{1}
{{4\pi n}}\left( {e^{i\frac{{2\pi n}}
{\ell }\left( {x - y} \right)}  - e^{ - i\frac{{2\pi n}}
{\ell }\left( {x - y} \right)} } \right)}  \hfill \\
\quad \quad \quad \quad \quad \quad \,{\kern 1pt} \, = i\frac{{x -
y}}
{{2\ell }} + \,\frac{1}
{{4\pi }}\ln \left( {1 - e^{ - i\frac{{2\pi }}
{\ell }\left( {x - y} \right)} } \right) - \frac{1}
{{4\pi }}\ln \left( {1 - e^{i\frac{{2\pi }}
{\ell }\left( {x - y} \right)} } \right) \hfill 
\end{gathered}
\end{equation} 
where we have used the expansion of the logarithm
\begin{equation} \sum\limits_{n = 1}^\infty  {\frac{{e^{ - \alpha n} }}
{n}}  =  - \ln (1 - e^{ - \alpha }
).\label{log} \end{equation} 
Since
\begin{displaymath} 
\ln \left( {\frac{{1 - e^{ - i\alpha } }}
{{1 - e^{i\alpha } }}} \right) = \left\{ \begin{gathered}
i\left( {\pi  - \alpha } \right){\text{    for }}\alpha  > 0 \hfill
\\
0{\text{               for }}\alpha {\text{ = 0}} \hfill \\
i\left( { - \pi  - \alpha } \right){\text{ for }}\alpha  < 0 \hfill \\
\end{gathered}  \right.
\end{displaymath} 
we have
\begin{equation} \left[ {\phi _R \left( x \right),\phi _R \left( y
\right)} \right] = \frac{i}
{4}\operatorname{sign} (x - y) = \left\{ \begin{gathered}
\frac{i}
{4}{\text{        for }}x - y > 0{\text{  }} \hfill \\
0{\text{         for  }}x - y = 0 \hfill \\
- \frac{i}
{4}{\text{     for  }}x - y < 0 \hfill 
\end{gathered}  \right.\label{phiLRcomm} \end{equation} 
Inserting this in equation (\ref{C})  we get
\begin{equation} \left\{ {\psi _R^\dagger  (x),\psi _R (y)} \right\} = \left\{
\begin{gathered}
0{\text{        for }}x \ne y{\text{ }} \hfill \\
2C^2 {\text{   for  }}x = y \hfill \\
\end{gathered}  \right.\end{equation} 
This expression agrees with (\ref{anticomm})  if the proportionality constant
\begin{math} C^2  = \delta (0)/2\end{math} is chosen to be a delta
function infinity (or alternatively is cutoff dependent, depending on
the summation range in (\ref{anticomm})).

Following the same steps for left movers, we find analogously 
\begin{equation} \left[ {\phi _L \left( x \right),\phi _L \left( y
\right)} \right] =  - \frac{i}
{4}\operatorname{sign} (x -
y) \label{e58} \end{equation} 

Similarly, the anti-commutation relations between right- and left
movers can be guaranteed by demanding suitable commutation relations
between the zero mode creation operators
\begin{displaymath} 
\left[ {\phi_R^0 ,\phi _L^0 } \right] = \frac{i}{4}
\end{displaymath} 
which yields 
\begin{math} \left[ {\phi _R^{} (x),\phi _L^{} (y)}
\right] = \frac{i}{4}\end{math}  
and therefore \begin{math}\left\{ {\psi _R  (x),\psi_L (y)} \right\} = 0\end{math}  
according to equations (\ref{phiLR})  and 
(\ref{bosR}).
Alternatively, the commutation relations between left- and right-movers
can also be fixed by using so-called Klein factors, which can also
handle several fermion channels.

\subsubsection{{Bosonic excitation and zero modes}}

At this point it is useful to pause and look at the excitations of the
bosonic Hilbert space in more detail.  For the bosonic
\textquotedblleft{}oscillator\textquotedblright{} modes it is clear
that the ground states of the Hamiltonian in (\ref{bosH2}) simply corresponds to the
vacuum \begin{math} \left| 0 \right\rangle \end{math} which can be
excited with an arbitrary number of bosons with the creation operators
\begin{math} b_n^{R \dagger } \end{math} and \begin{math}b_n^{L \dagger} \end{math}
for each \begin{math} n\end{math} separately.  The vacuum is defined as
\begin{equation}
b_n|0\rangle = 0 \ \ \ \ \rm for\  all\  \it n \label{vacuum}
\end{equation}
The total fermion number relative to the ground state can be added and
removed with the zero mode operators \begin{math} \exp \left( { - i\sqrt
{4\pi } \phi _0^R } \right)\end{math}  and \begin{math}\exp \left(
{i\sqrt {4\pi } \phi _0^R } \right)\end{math}, respectively.  The
corresponding relation \begin{math} n_R e^{ - i\sqrt {4\pi } \phi
_0^R } \left| \lambda  \right\rangle  = e^{ - i\sqrt {4\pi } \phi
_0^R } \left( {n_R  + 1} \right)\left| \lambda  \right\rangle
\end{math}  for any state \begin{math}\left| \lambda  \right\rangle
\end{math}  can be shown with the help of equation (\ref{e49})  and is left as an
exercise here (remember that \begin{math} n_R  = Q_R /\sqrt \pi 
\end{math}  is the counting operator).  The fermion number creation
always acts from below, i.e. the entire corresponding state is
\textquotedblleft{}pushed up\textquotedblright{} one step by acting
with \begin{math} \exp \left( { - i\sqrt {4\pi } \phi _0^R }
\right)\end{math}  including any possible particle-hole excitations.  An
arbitrary number of fermions can be removed or added, but only in
integer numbers (see also Fig.~\ref{f16}).  The mathematical reason for
this is actually the periodic boundary condition \begin{math} \psi (x_j
) = \psi (x_{j + N} )\end{math} .  According to equations (\ref{psi}) and (\ref{psiLR})  the boundary
condition implies for the left and right moving fields
\begin{equation} \psi _R (x) = \psi _R (x + \ell )e^{ik_F N} ,  
\hspace{15pt}  \psi _L (x) = \psi _L (x + \ell )e^{ - ik_F
N} \end{equation} 
This translates into conditions on the boson field (\ref{phiLR}) by the use of the
bosonization formula (\ref{bosR}).  Since the sum of oscillator modes in the boson
field operator \begin{math} \sum\limits_{n = 1}^\infty  {\frac{1}
{{\sqrt {4\pi n} }}} (e^{i\frac{{2\pi n}}
{\ell }x} b_n^R  + e^{ - i\frac{{2\pi n}}
{\ell }x} b_n^{R \dagger } )\end{math}  is already periodic in
\begin{math} \ell \end{math}, we are left with the following condition
on the zero modes (using  (\ref{baker}) since \begin{math} \phi _0^R \end{math} and
\begin{math} Q_R \end{math} do not commute):
\begin{eqnarray}
& & e^{i\sqrt{4 \pi}\phi_0^R} = e^{i\sqrt{4 \pi }(\phi_0^R+Q_R)}e^{ik_FN} = 
e^{i\sqrt{4 \pi}\phi_0^R} e^{i\sqrt{4 \pi}Q_R} e^{4\pi [\phi_0^R,Q_R]/2} e^{ik_FN} \nonumber \\
& \longrightarrow &  1 = e^{i\sqrt{4 \pi}Q_R} e^{-i \pi } e^{ik_FN} \nonumber \\
& \longrightarrow & \sqrt{4 \pi} Q_R = 2 \pi n - k_FN+\pi \label{e60}
\end{eqnarray}
We see that the number operator \begin{math} Q_R /\sqrt \pi   = n_R
\end{math}  must indeed be an integer, since \begin{math}k_F N = (n_0  +
1)\pi \end{math}  is an odd multiple of
\textit{\begin{math} \pi{}\end{math}} according to equation (\ref{n0}). 
Therefore, zero mode particle excitations  \begin{math} \exp \left( {
- i\sqrt {4\pi } \phi _0^R } \right)\end{math}  can only be
added/removed in integral numbers.  Arbitrary real numbers in the
exponent would lead to unphysical states outside the Hilbert space. 
Interestingly, the spectrum and Hilbert-space of the zero modes
therefore have a one-to-one correspondence to a single particle on a
ring, where \textit{Q$_{R}$} plays the role of the momentum operator
and \begin{math} \phi _0^R \end{math} is the corresponding position
operator with canonical commutation relations as defined in (\ref{zerocomm}).

\subsubsection{{Summary of the bosonization formulas}}

Before we continue, let us summarize the most important bosonization
formulas.

~\\ Linearization

\begin{math} c_k^R  = c_{k_F  + k}
\end{math} \hspace{15pt}\hspace{15pt}\hspace{15pt}\hspace{15pt}\hspace{15pt}\hspace{15pt}\begin{math}c_k^L
 = c_{ - k_F  + k}
\end{math} \hfill (\ref{e13})

\begin{math} H = \sum\limits_k^{} {v_F k} \left( {c_k^{R \dagger } c_k^R  -
c_k^{L \dagger} c_k^L }
\right)\end{math} \hfill (\ref{linH})

\begin{math} \psi (x_j ) = \sqrt a \left( {e^{ik_F x_j /a} \psi _R (x_j
) + e^{ - ik_F x_j /a} \psi _L (x_j )}
\right)\end{math} \hfill (\ref{psi})

\begin{math} \psi _R (x_j ) = \frac{1}
{{\sqrt \ell  }}\sum\limits_{k =  - \infty }^\infty  {c_k^R e^{ikx_j
/a} }
\end{math} \hspace{15pt}\hspace{15pt}\hspace{15pt}\hspace{15pt}\begin{math}\psi
_L (x_j ) = \frac{1}
{{\sqrt \ell  }}\sum\limits_{k =  - \infty }^\infty  {c_k^L e^{ikx_j
/a} } \end{math} \hfill (\ref{psiLR})

~\\ Bosons in terms of fermions

\begin{math}  - i\sqrt {\frac{{kN}}
{{2\pi }}} b_k^{R \dagger }  = \rho _k^R  = \int_0^\ell  {:\rho _R
(x):e^{ikx/a} dx}  = \sum\limits_{k'} {c_{k' + k}^{R \dagger } c_{k'}^R }
\end{math} \hspace{15pt} for
$k>0$ \hfill (\ref{bR})

\begin{math} i\sqrt {\frac{{kN}}
{{2\pi }}} b_k^{L \dagger}  = \rho _{ - k}^L  = \int_0^\ell  {:\rho _L
(x):e^{ - ikx/a} dx}  = \sum\limits_{k'} {c_{k' - k}^{L \dagger} c_{k'}^L }
\end{math}  \hspace{15pt} for
$k>0$ \hfill (\ref{bL})

\begin{math} Q_{R/L} /\sqrt \pi   = n_{R/L}  = \int_0^\ell  {:\rho
_{R/L} (x):dx} \end{math} \hfill (\ref{intrho0})   

\begin{math} H = v_F \sum\limits_{k > 0} {k\left( {b_k^{R \dagger } b_k^R  +
b_k^{L \dagger} b_k^L } \right)}  + \frac{{\pi v_F }}
{N}\left( {n_R^2  + n_L^2 }
\right)\end{math} \hfill (\ref{bosH})

~\\Fermions in terms of bosons fields

\begin{math} :\rho _R (x):\; = \frac{1}
{{\sqrt \pi  }}\partial _x \phi _R (x)\quad \quad \quad \quad \quad
:\rho _L (x):\; = \frac{1}
{{\sqrt \pi  }}\partial _x \phi _L
(x)\end{math} \hfill (\ref{rhoLR})

\begin{math} \psi _R (x) \propto \exp \left( {i\sqrt {4\pi } \phi _R
(x)} \right)\end{math} \hspace{15pt} \hspace{15pt}\begin{math}\psi _L
(x) \propto \exp \left( { - i\sqrt {4\pi } \phi _L (x)}
\right)\end{math} \hfill (\ref{bosR})

\begin{math} H = av_F \int_0^\ell  {dx\left( {(\partial _x \phi _R
)^2  + (\partial _x \phi _L )^2 } \right)}
\end{math} \hfill (\ref{bosH3})

where

\begin{math} \begin{gathered}
\phi _R (x) = \phi _0^R  + Q_R \frac{x}
{\ell } + \sum\limits_{n = 1}^\infty  {\frac{1}
{{\sqrt {4\pi n} }}} \left( {e^{i\frac{{2\pi n}}
{\ell }x} b_n^R  + e^{ - i\frac{{2\pi n}}
{\ell }x} b_n^{R \dagger } } \right) \hfill \\
\phi _L (x) = \phi _0^L  + Q_L \frac{x}
{\ell } + \sum\limits_{n = 1}^\infty  {\frac{1}
{{\sqrt {4\pi n} }}} \left( {e^{ - i\frac{{2\pi n}}
{\ell }x} b_n^L  + e^{i\frac{{2\pi n}}
{\ell }x} b_n^{L \dagger} } \right) \hfill \\
\end{gathered}
\end{math} \hfill (\ref{phiLR})

\subsection{Correlation functions}

As a typical application for bosonization, we would now like to
calculate the correlation function for the fermion fields.

\subsubsection{In space}

Correlations along the wire decay according to the following
expression
\begin{equation} \left\langle {\psi ^\dagger  (x)\psi (y)} \right\rangle 
\approx ae^{ - ik_F \left( {x - y} \right)/a} \left\langle {\psi _R^\dagger 
(x)\psi _R (y)} \right\rangle  + ae^{ik_F \left( {x - y} \right)/a}
\left\langle {\psi _L^\dagger  (x)\psi _L (y)} \right\rangle
\label{e61}
\end{equation} 
in the linearized approximation (\ref{psi}).  The left and right movers are
uncorrelated \begin{displaymath} \left\langle {\psi _L^\dagger  (x)\psi _R (y)}
\right\rangle  = \left\langle {\psi _R^\dagger  (x)\psi _L (y)}
\right\rangle  = 0,\end{displaymath}  because of the zero modes:  If a
right-mover is added \begin{math} \exp \left( { - i\sqrt {4\pi } \phi
_0^R } \right)\end{math}  it must also be removed again with
\begin{math} \exp \left( {i\sqrt {4\pi } \phi _0^R }
\right)\end{math}  in order to get the ground state back.  Therefore, in
particular \begin{math} \left\langle 0 \right|\exp \left( { - i\sqrt
{4\pi } \phi _0^R } \right)\left| 0 \right\rangle  = 0\end{math} .

Using the bosonization formula  in real space we can write for the
right moving correlation function (also using (\ref{baker})  and (\ref{phiLRcomm}))
\begin{equation} \begin{gathered}
\left\langle {\psi _R^\dagger  \left( x \right)\psi _R \left( y \right)}
\right\rangle  = C^2 \left\langle {e^{ - i\sqrt {4\pi } \left( {\phi
_R \left( x \right) - \phi _R \left( y \right)} \right)} }
\right\rangle   e^{2\pi \left[ {\phi _R \left( x
\right),\phi _R \left( y \right)} \right]}  \hfill \\
\quad \quad \quad \quad \quad \quad \,\,{\kern 1pt} {\kern 1pt}  =
C^2 \left\langle {e^{ - i\sqrt {4\pi } \left( {\phi _R \left( x
\right) - \phi _R \left( y \right)} \right)} } \right\rangle  
e^{i\pi {\text{sign}}(x - y)/2}  \label{e62}
\end{gathered}
\end{equation} 
In order to calculate the expectation value of the exponential, we have
to consider the zero modes and the oscillator modes separately.  The
expectation values of the zero modes are simple since the
addition/removal operators cancel in the sum
\textit{\begin{math} \phi{}\end{math}$_{R}(x)-\phi{}_{R}(y)$}
and we are left with
\begin{equation} \left\langle 0 \right|\exp \left( { - i\sqrt {4\pi }
Q_R^{} (x - y)/\ell } \right)\left| 0 \right\rangle  =
1\end{equation}
which follows from the fact that the ground state has no additional
particles \begin{math} Q_R \left| 0 \right\rangle  = 0\end{math}.

For the ground state expectation value of the oscillator modes we can
use the Baker-Campbell-Hausdorff formula (\ref{baker})  in order to arrive at the general
expression
\begin{equation} \left\langle {0\left| {\exp (\alpha b + \beta b^\dagger  )}
\right|0} \right\rangle  = \left\langle 0 \right|e^{\alpha b^\dagger  }
e^{\beta b} \left| 0 \right\rangle e^{\frac{{\alpha \beta }}
{2}\left[ {b,b^\dagger  } \right]}  = \exp (\alpha \beta
/2) \label{e63}\end{equation} 
where we have also used equation (\ref{vacuum}).  In fact the relation (\ref{e63})  is just a
special case for \begin{math} T=0\end{math} of the more general
cummulant theorem for bosons \begin{math} \left\langle {\exp (f)}
\right\rangle  = \exp \left\langle {f^2 } \right\rangle /2{\text{
}}\end{math} (see appendix equation (\ref{a1})), which is valid for any
temperature and any linear combination of bosons
\begin{math} f\end{math} as shown in the appendix.  With the help of
the cummulant formula (\ref{a1}) it is also possible to calculate correlation
functions at finite temperature and finite system sizes, which is
explained in detail in Ref.~\cite{15}
(also for \textquotedblleft{}open\textquotedblright{} boundary
conditions).  For simplicity, we will restrict ourselves to the ground
state correlation at \begin{math} T=0\end{math}.

Equation  (\ref{e62}) can be evaluated by inserting the mode expansion  (\ref{phiLR}) into
(\ref{e63})
\begin{equation} \left\langle {\psi _R^\dagger  \left( x \right)\psi _R \left( y
\right)} \right\rangle  = iC^2 \,{\text{sign}}(x - y)\left\langle e^{
 { - \sum\limits_n {\frac{1}
{{2n}}\left( {e^{i\frac{{2\pi }}
{\ell }nx}  - e^{i\frac{{2\pi }}
{\ell }ny} } \right)\left( {e^{ - i\frac{{2\pi }}
{\ell }nx}  - e^{ - i\frac{{2\pi }}
{\ell }ny} } \right)} } } \right\rangle \end{equation} 
Using again the expansion of the logarithm (\ref{log}),  we find for the sum
\begin{displaymath} 
\begin{gathered}
\sum\limits_{n > 0} {\frac{1}
{n}\left( {e^{i\frac{{2\pi }}
{\ell }nx}  - e^{i\frac{{2\pi }}
{\ell }ny} } \right)\left( {e^{ - i\frac{{2\pi }}
{\ell }nx}  - e^{ - i\frac{{2\pi }}
{\ell }ny} } \right)}  \hfill \\
\quad \quad \quad \quad \quad \quad \quad \quad \quad \quad \quad
\quad \quad  = \ln \left( {1 - e^{i\frac{{2\pi }}
{\ell }(x - y)} } \right) + \ln \left( {1 - e^{ - i\frac{{2\pi }}
{\ell }(x - y)} } \right) - 2\mathop {\lim }\limits_{\varepsilon  \to
0} \left( {\ln \varepsilon } \right) \hfill \\
\end{gathered}
\end{displaymath} 
where the last term corresponds again to an infinity
\begin{equation} \sum\limits_{n = 1}^\infty  {\frac{1}
{n}}  = \mathop {\lim }\limits_{\varepsilon  \to 0} \sum\limits_{n =
1}^\infty  {\frac{{e^{ - \varepsilon } }}
{n}}  =  - \mathop {\lim }\limits_{\varepsilon  \to 0} \ln \left( {1 -
e^{ - \varepsilon } } \right) =  - \mathop {\lim }\limits_{\varepsilon 
\to 0} \ln \varepsilon.
\end{equation} 
However, this infinity combines with the prefactor \textit{$C^{2}$}, so
that a finite result is obtained using the relation 
\textit{\begin{math} \epsilon{}\end{math}$C^{2
}$=}1\textit{/\begin{math} \ell\end{math}. }
\begin{equation} \begin{gathered}
\left\langle {\psi _R^\dagger  \left( x \right)\psi _R \left( y \right)}
\right\rangle  = \mathop {\lim }\limits_{\varepsilon  \to 0} iC^2
\,{\text{sign}}(x - y)\exp \left( {\frac{1}
{2}\ln \frac{\varepsilon }
{{1 - e^{i\frac{{2\pi }}
{\ell }(x - y)} }}\frac{\varepsilon }
{{1 - e^{ - i\frac{{2\pi }}
{\ell }(x - y)} }}} \right) \hfill \\
\quad \quad \quad \quad \quad \quad \;\; = \frac{i}
{{2\ell }}\frac{{\,{\text{sign}}(x - y)}}
{{\left| {\sin \frac{\pi }
{\ell }(x - y)} \right|}} \hfill \\
\quad \quad \quad \quad \quad \quad \;\; = \frac{{i\,}}
{{2\ell \sin \frac{\pi }
{\ell }(x - y)}} \hfill \\
\end{gathered} \label{e65} \end{equation} 
The relation \textit{\begin{math} \epsilon{}\end{math}$C^{2
}$=}1\textit{/\begin{math} \ell\end{math}} can be shown, e.g.~by
evaluating the correlation function (\ref{e62}) directly from the definition (\ref{psiLR})
by using the geometrical sum (and the quantization of the
\begin{math} k\end{math}-values relative to \textit{k$_{F}$} as in
equation (\ref{e20})).  The calculation for the left moving correlation function
follows the same steps, except for a different sign in equation (\ref{e58})
\begin{equation} \left\langle {\psi _L^\dagger  \left( x \right)\psi _L \left( y
\right)} \right\rangle  =  - \frac{{i\,}}
{{2\ell \sin \frac{\pi }
{\ell }(x -
y)}}\end{equation} 
Therefore, in summary from equation (\ref{e61})
\begin{equation} \left\langle {\psi ^\dagger  (x)\psi (y)} \right\rangle  =
\frac{{\sin k_F (x - y)\,}}
{{N\sin \frac{\pi }
{\ell }(x -
y)}}\end{equation} 
It is left as an exercise for the reader to show that this is indeed
the exact correlation function which can also be obtained without any
approximations directly from (\ref{e3}).

\subsubsection{In time}

The time correlation function plays an important role, e.g.~in order
to evaluate the Green's function (\ref{G}).  Before we can evaluate the
expectation values we have to determine how the operators evolve in
time.  Using the general formula
\begin{equation} A(t) = e^{iHt} Ae^{ - iHt}
\end{equation} 
it is straightforward to show that
\begin{equation} \begin{gathered}
b_n^{R/L} (t)\;\,{\kern 1pt}  = b_n^{R/L} e^{ - in\frac{{2\pi v_F }}
{N}t}  \hfill \\
b_n^{R/L \dagger} (t) = b_n^{R/L \dagger} e^{in\frac{{2\pi v_F }}
{N}t}  \hfill \\
Q_{R/L} (t)\;\, = Q_{R/L}  \hfill \\
\phi _0^{R/L} (t)\,\, = \phi _0^{R/L}  \mp Q_{R/L} \frac{{v_F
}}
{N}t \hfill \\
\end{gathered}
\end{equation} 
Inserting these results into equation (\ref{phiLR})  we can immediately obtain a
time-dependent mode expansion
\begin{equation} \begin{gathered}
\phi _R (x,t) = \phi _0^R  + Q_R \frac{{x - av_F t}}
{\ell } + \sum\limits_{n = 1}^\infty  {\frac{1}
{{\sqrt {4\pi n} }}} \left( {e^{i\frac{{2\pi n}}
{\ell }(x - av_F t)} b_n^R  + e^{ - i\frac{{2\pi n}}
{\ell }(x - av_F t)} b_n^{R \dagger } } \right) \hfill \\
\phi _L (x,t) = \phi _0^L  + Q_L \frac{{x + av_F t}}
{\ell } + \sum\limits_{n = 1}^\infty  {\frac{1}
{{\sqrt {4\pi n} }}} \left( {e^{ - i\frac{{2\pi n}}
{\ell }(x + av_F t)} b_n^L  + e^{i\frac{{2\pi n}}
{\ell }(x + av_F t)} b_n^{L \dagger} } \right) \hfill \\
\end{gathered} \end{equation} 
This is a remarkable result in many ways.  First of all we see that
the right- and left-movers are only functions of the right and
left-moving light-cone coordinates
\textit{x\begin{math} \pm{}\end{math}av$_{F}$t}.  Secondly, with the
help of this result the calculation of any time-space correlation
function is no more or less difficult than the calculation of a pure
space correlation function by substituting the time dependent light
cone coordinates for \begin{math} x-y\end{math} in (\ref{e65}).  That means that we
have in principle the tools to solve any dynamic problem.  In
particular, we can immediately generalize the result (\ref{e65}) to express the
time correlation function
\begin{equation} \left\langle {\psi _R^\dagger  (t)\psi _R (0)} \right\rangle  =
 - \frac{{i\,}}
{{2\ell \sin \frac{\pi }
{N}v_F
t}}\end{equation} 
where it must be remembered that \begin{math} t = \mathop {\lim
}\limits_{\varepsilon  \to 0} \left( {t - i\varepsilon }
\right)\end{math}  in order to ensure that the summation converges.

\subsection{Spin-charge separation }

One of the most famous and surprising results of one-dimensional
many-body physics is the prediction of separate spin- and
charge-excitations.  We will show here that on the level of
bosonization the spin-charge separation simply corresponds to taking
suitable linear combinations of degenerate states.

\subsubsection{Spin and charge excitations }

All calculations that we have done so far actually also apply for
electrons with spin if we use two species of bosons
$\phi_\sigma$
with index
$\sigma=\downarrow, \uparrow$,
one for spin up and one for spin down.  According to equation (\ref{rhoLR}) the bosons can
immediately be related to fermionic densities with spin up and spin
down.
\begin{equation} 
:\rho _{R/L}^\sigma  (x):\; = \frac{1}
{{\sqrt \pi  }}\partial _x \phi _{R/L}^\sigma  (x)\quad \quad \quad
\quad \quad \sigma  =  \uparrow , \downarrow
\end{equation} 
If we are interested in the total charge density at a point
\begin{math} x\end{math}, we will use the sum of spin up and spin down
\begin{equation} :\rho _{R/L}^c (x):\; = :\rho _{R/L}^ \uparrow  (x): +
:\rho _{R/L}^ \downarrow 
(x):\end{equation} 
Likewise, we can calculate the total spin density at point
\begin{math} x\end{math} by using the difference
\begin{equation} :\rho _{R/L}^s (x):\; = :\rho _{R/L}^ \uparrow  (x): -
:\rho _{R/L}^ \downarrow 
(x):\end{equation} 
Analogously, we can define new boson fields that correspond to the spin
and charge densities
\begin{equation} \begin{gathered}
\phi _{R/L}^c (x) = \frac{1}
{{\sqrt 2 }}\left( {\phi _{R/L}^ \uparrow  (x) + \phi _{R/L}^
\downarrow  (x)} \right) \hfill \\
\phi _{R/L}^s (x) = \frac{1}
{{\sqrt 2 }}\left( {\phi _{R/L}^ \uparrow  (x) - \phi _{R/L}^
\downarrow  (x)} \right) \hfill \\
\end{gathered}
\label{e72}
\end{equation} 
This defines new zero modes and oscillator mode operators (the index
for left and right movers \begin{math} L/R\end{math} is omitted here)
\begin{equation} \begin{gathered}
b_n^{c/s}  = \frac{1}
{{\sqrt 2 }}\left( {b_n^ \uparrow  \pm b_n^ \downarrow  }
\right) \hfill \\
\phi _{0}^{c/s}  = \frac{1}
{{\sqrt 2 }}\left( {\phi _{0}^{\uparrow }  \pm \phi _{0}^{
\downarrow } } \right) \hfill \\
Q_{}^{c/s}  = \frac{1}
{{\sqrt 2 }}\left( {Q_{}^ \uparrow   \pm Q_{}^ \downarrow}  \right) \hfill \\
\end{gathered}
\label{e73}
\end{equation} 
which have again canonical commutation 
relations analogous to (\ref{e29}) and (\ref{zerocomm})  
as can easily be verified.  The mode expansion of the fields (\ref{e72})  in terms of
the new operators (\ref{e73}) is then the same as before.

It is a straightforward exercise to write the bosonization formulas (\ref{rhoLR})
and  (\ref{bosR}) in terms of the new spin and charge fields, for example:
\begin{equation} \begin{gathered}
\psi _R^ \uparrow  (x) \propto \exp \left( {i\sqrt {4\pi } \phi
_R^ \uparrow  (x)} \right) = \exp i\sqrt {2\pi } \left( {\phi _R^c
(x) + \phi _R^s (x)} \right) \hfill \\
:\rho _R^ \downarrow  (x):\; = \frac{1}
{{\sqrt \pi  }}\partial _x \phi _R^ \downarrow  (x) = \frac{1}
{{\sqrt {2\pi } }}\left( {\partial _x \phi _R^c (x) - \partial _x
\phi _R^s (x)} \right) \hfill \\
:\rho _R^{s/c} (x):\; = \sqrt {\frac{2}
{\pi }} \partial _x \phi _R^{s/c} (x)\quad \quad \quad  \hfill \\
\end{gathered} \label{e74}\end{equation} 
Most importantly, the Hamiltonian (\ref{bosH3})  can also be written in terms of the
new spin and charge operators
\begin{equation} \begin{gathered}
H = av_F \int_0^\ell  {dx\left( {(\partial _x \phi _R^ \uparrow 
)^2  + (\partial _x \phi _R^ \downarrow  )^2  + (\partial _x \phi
_L^ \uparrow  )^2  + (\partial _x \phi _L^ \downarrow  )^2 }
\right)}  \hfill \\
\quad {\kern 1pt} {\kern 1pt}  = av_F \int_0^\ell  {dx\left(
{(\partial _x \phi _R^c )^2  + (\partial _x \phi _R^s )^2  +
(\partial _x \phi _L^c )^2  + (\partial _x \phi _L^s )^2 }
\right)}  \hfill \\
\end{gathered}  \label{e75}\end{equation} 

Since the spin and charge excitations appear separately in the
Hamiltonian (\ref{e75}), the partition function factorizes.  It is therefore
possible to regard the spin and charge particles as the
\textquotedblleft{}new\textquotedblright{} independent fundamental
excitations instead of the \textquotedblleft{}old\textquotedblright{}
spin up and spin down particles.  Indeed, because of the degeneracy of
the spin up and spin down channel, we could have used any canonical
rotation to define new particles, but spin and charge are particularly
useful when interactions are present.  This is intuitively clear
because realistic interactions will couple total charge densities and
thereby lift the degeneracy between the spin and the charge channel. 
With interactions we are therefore forced to use the picture of spin
and charge excitations, since the freedom of rotating degenerate
channels is lost.

It should be noted that the spin and charge separation is not obeyed
exactly for the particle numbers, because the
\textquotedblleft{}new\textquotedblright{} zero modes still must obey
the \textquotedblleft{}old\textquotedblright{} quantization formula (\ref{e60}). 
In particular, the number of charge and spin particles is given by
\begin{equation} \begin{gathered}
n_R^c  = n_R^ \uparrow   + n_R^ \downarrow   = \sqrt {\frac{2}
{\pi }} Q_R^c  \hfill \\
n_R^s  = n_R^ \uparrow   - n_R^ \downarrow   = \sqrt {\frac{2}
{\pi }} Q_R^s  \hfill \\
\end{gathered}
\end{equation} 
Both numbers must be integers, but cannot be changed independently. 
Adding a spin particle must always be accompanied by adding or removing
a charge particle and vice versa, i.e.~the total sum of spin and charge
particles must always remain even.  This just reflects the fact that we
always have to add and remove real electrons, instead of spin/charge
quasi-particles.

\begin{figure}
\begin{center}
\includegraphics[width=350pt]{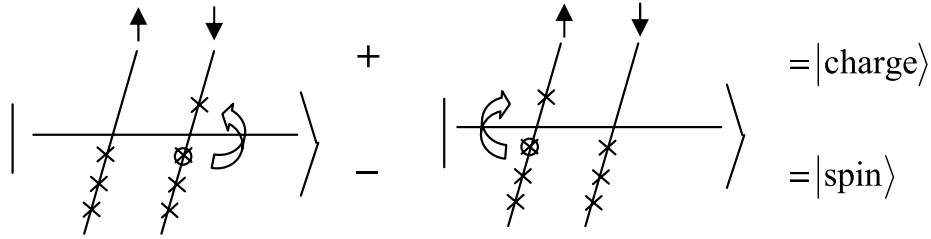}
\caption{Charge excitations are  symmetric linear combinations of up and down excitations.
Spin excitations are anti-symmetric linear combinations of up and down.}
\label{f17}
\end{center}
\end{figure}

We are now in the position to also interpret the nature of the spin
and charge excitations.  As discussed in section 2.2 we understand
the bosonic excitations in terms of fermions being shifted up the
spectrum.  The spin and charge excitation now simply correspond to odd
and even linear combinations of those shifts according to equation (\ref{e73}). 
This is depicted schematically in diagram \ref{f17}.

\subsubsection{Spin and charge correlation functions }

We now have the tools in order to calculate correlation functions of
electrons with spin
\begin{displaymath} 
\left\langle {\psi ^{ \uparrow  \dagger } (x)\psi ^ \uparrow  (y)}
\right\rangle  \approx ae^{ - ik_F \left( {x - y} \right)} \left\langle
{\psi _R^{ \uparrow \dagger  } (x)\psi _R^ \uparrow  (y)} \right\rangle  +
ae^{ik_F \left( {x - y} \right)} \left\langle {\psi _L^{ \uparrow  \dagger }
(x)\psi _L^ \uparrow  (y)} \right\rangle
\end{displaymath} 
Using equation (\ref{e74})  we can calculate the right moving part (using the same
calculations as in section 2.5)
\begin{equation} \begin{gathered}
\left\langle {\psi _R^{ \uparrow  \dagger } \left( x \right)\psi _R^
\uparrow  \left( y \right)} \right\rangle  \propto \,{\kern 1pt} {\kern
1pt} i{\text{ sign}}\left( {x - y} \right)\left\langle {e^{ - i\sqrt
{4\pi } \left( {\phi _R^ \uparrow  \left( x \right) - \phi _R^
\uparrow  \left( y \right)} \right)} } \right\rangle  \hfill \\
\quad \quad \quad \quad \quad \quad \;\; = i{\text{ sign}}\left( {x
- y} \right)\left\langle {e^{ - i\sqrt {2\pi } \left( {\phi _R^c
\left( x \right) - \phi _R^c \left( y \right)} \right)} }
\right\rangle  \times \left\langle {e^{ - i\sqrt {2\pi } \left(
{\phi _R^s \left( x \right) - \phi _R^s \left( y \right)}
\right)} } \right\rangle  \hfill \\
\quad \quad \quad \quad \quad \quad \;\; = i{\text{ sign}}\left( {x
- y} \right)\frac{1}
{{\sqrt {\left| {\sin \frac{{\pi \left( {x - y} \right)}}
{\ell }} \right|} }} \times \frac{1}
{{\sqrt {\left| {\sin \frac{{\pi \left( {x - y} \right)}}
{\ell }} \right|} }} \hfill \\
\end{gathered} \label{e76} \end{equation} 
Without interactions the outcome is of course identical to the
spinless result (\ref{e65}), but the calculation shows that the correlation
function factorizes into a spin and a charge part.  When interactions
are introduced in the next section we expect that the degeneracy of
spin and charge is lifted and therefore the correlations decay
differently and independently for spin- and charge-like excitations.

\section{Electron-electron interactions}

Finally we are now in the position to deal with interaction effects,
which of course is the main goal of bosonization.  After having derived
the mathematical foundation of bosonization in the previous section, it
is now straightforward to apply that prescription to typical
interaction terms discussed above (see equations (\ref{e5})  and (\ref{e6}))

\subsection{Scattering processes}

\begin{figure}
\begin{center}
\includegraphics[width=378pt]{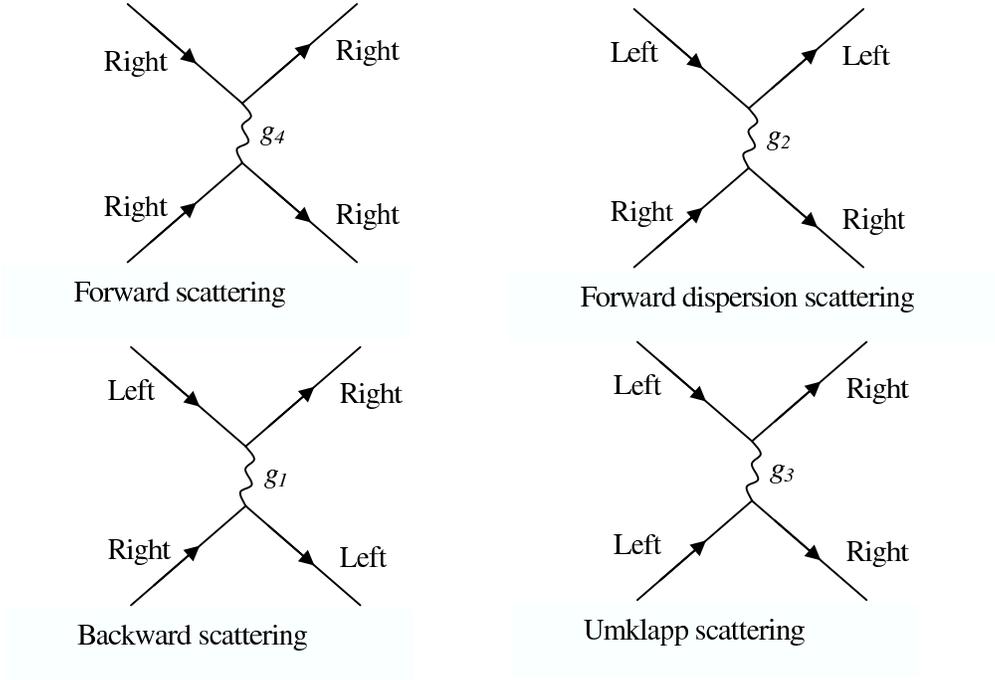}
\caption{The different scattering processes in the linearized approximation.}
\label{f18}
\end{center}
\end{figure}

Let us consider the standard model of density-density interactions in
equation (\ref{e5})
\begin{displaymath} 
H_{\operatorname{int} }  = \sum\limits_{j = 1}^N {\sum\limits_{m =
1}^N {\psi ^\dagger  (x_j )\psi (x_j )U(m)\psi ^\dagger  (x_{j + m} )\psi (x_{j +
m} )} }
\end{displaymath} 
or equivalently in terms of the wave-numbers \begin{math} k\end{math}
in equation (\ref{e6})
\begin{equation} H_{\operatorname{int} }  = \frac{1}
{N}\sum\limits_{k,k',\Delta k} {c_k^\dagger  c_{k - \Delta k} U(\Delta
k)c_{k'}^\dagger  } c_{k' + \Delta k}
.\end{equation} 
Following the standard program of bosonization, the first step is
always to restrict the creation and annihilation operators in a
linearized region
\textit{\begin{math} \vert{}\end{math}k\textendash{}k$_{F}$\begin{math}\vert{}\end{math}\begin{math}<\end{math}\begin{math}\Lambda{}\end{math}}
and
\textit{\begin{math} \vert{}\end{math}k+k$_{F}$\begin{math}\vert{}\end{math}\begin{math}<\end{math}\begin{math}\Lambda{}\end{math}}
around \textit{k$_{F}$} according to Fig.~\ref{f14} (see section 2.1). 
Let us systematically analyze the different possibilities as shown in Fig.~\ref{f18}.

{\bf ~\\Forward scattering}

We first consider the case of small momentum transfer
\textit{\begin{math} \Delta{}k<\Lambda{}\end{math}},
which is commonly referred to as \textit{forward scattering}, since a
right mover is always scattered into a right mover and a left mover is
always scattered into a left mover.  Using equation (\ref{e13})  and restricting
the momenta \begin{math} k\end{math} and \textit{k'} to left movers
$-k_{F}- \Lambda{}<k,\ k'<-k_{F}+\Lambda{}$
or right movers
$k_{F}- \Lambda{}<k,\ k'<k_{F}+\Lambda{}$,
the corresponding interaction Hamiltonian is given by
\begin{equation} \begin{gathered}
H_{{\text{forward}}}  = \frac{1}
{N}\sum\limits_{k,k',\Delta k} {\left( {c_k^{R \dagger } c_{k - \Delta k}^R 
+ c_{k'}^{L \dagger} c_{k' - \Delta k}^L } \right)U(\Delta k)\left(
{c_{k'}^{R \dagger } c_{k' + \Delta k}^R  + c_{k'}^{L \dagger} c_{k' + \Delta k}^L
} \right)}  \hfill \\
\quad {\kern 1pt} \quad {\kern 1pt} \;\,\; = \frac{1}
{N}\sum\limits_{\Delta k =  - \infty }^\infty  {\left( {g_2 \left(
{\rho _{ - \Delta k}^R \rho _{\Delta k}^L  + \rho _{ - \Delta k}^L \rho
_{\Delta k}^R } \right) + g_4 \left( {\rho _{ - \Delta k}^R \rho
_{\Delta k}^R  + \rho _{ - \Delta k}^L \rho _{\Delta k}^L } \right)}
\right)}  \hfill \\
\quad {\kern 1pt} \quad {\kern 1pt} \;\,\; = \frac{1}
{N}\sum\limits_{n = 1}^\infty  {\left( {2ng_2 \left( {b_n^{R \dagger }
b_n^{L \dagger}  + b_n^R b_n^L } \right) + 2ng_4 \left( {b_n^{R \dagger } b_n^R  +
b_n^{L \dagger} b_n^L } \right)} \right.}  \hfill \\
\quad {\kern 1pt} \quad {\kern 1pt} \;\,\;\quad {\kern 1pt} \quad
{\kern 1pt} \;\,\;\quad {\kern 1pt} \left. { + 2g_2 n_R^{} n_L^{}  +
g_4 \left( {n_R^2  + n_L^2 } \right) + const.} \right) \hfill \\
\end{gathered} \label{e77} \end{equation} 
where we have used equations (\ref{rhoR})  and (\ref{bL}).  The constant comes from ordering
the bosons and plays no significant role in the Hamiltonian.  It is
customary to denote the right-left interaction by
\textit{$g_{2}=$U\begin{math}( \Delta{}k)\end{math}}, which is sometimes
also called dispersion scattering.  The density-density interaction
strength on the same branch is denoted by
\textit{$g_{4}=$U\begin{math} (\Delta{}k)\end{math}}.  Often
\textit{$g_{2}$} and \textit{$g_{4}$} are taken to be independent of
momentum, which is justified for short range potentials
\begin{math} U(m) \end{math} that drop off quickly for
\textit{\begin{math} m>1/\Lambda{}\end{math}}.  In
that case, we can write the forward scattering term more compactly with
the help of equation (\ref{phiLR})
\begin{equation} H_{{\text{forward}}}  = \frac{a}
{\pi }\int_0^\ell  {dx\;\left( {2g_2 \left( {\partial _x \phi _R
\partial _x \phi _L } \right) + g_4 \left( {(\partial _x \phi _R
)^2  + (\partial _x \phi _L )^2 } \right)} \right)} \label{e78}
\end{equation} 

{~\\  \bf Backward scattering}

There are additional terms in equation (\ref{e6})  corresponding to large
momentum transfer around \begin{math} \pm{}\end{math}\textit{$2k_{F}$},
which scatter left into right movers and vice versa.  We use a new
parameterization \begin{math} \Delta k =  \pm \left( {2k_F  + \delta k}
\right)\end{math}, where \begin{math}\delta{}\end{math}k is now assumed
to be small.  Using (\ref{e6}) and (\ref{e13}), we obtain
\begin{displaymath} 
\begin{gathered}
H_{{\text{backward}}}  = \frac{1}
{N}\sum\limits_{k,k',\delta k} {c_k^{R \dagger } c_{k - \delta k}^L U(2k_F 
+ \delta k)c_{k'}^{L \dagger} c_{k' + \delta k}^R  + c_k^{L \dagger} c_{k -
\delta k}^R U( - 2k_F  - \delta k)c_{k'}^{R \dagger } c_{k' + \delta k}^L } 
\hfill \\
\quad {\kern 1pt} \quad {\kern 1pt} \;\;\,\; = \frac{1}
{N}\sum\limits_{k,k',\Delta k} {g_1 \left( {c_k^{R \dagger } c_{k + \Delta
k}^R c_{k' - \Delta k}^L c_{k'}^{L \dagger}  + c_k^{L \dagger} c_{k + \Delta k}^L
c_{k' - \Delta k}^R c_{k'}^{R \dagger } } \right)}  \hfill \\
\end{gathered}
\end{displaymath} 
where we have again re-parameterized \begin{math} \Delta k = \delta k +
k' - k\end{math}  in the last step and defined the backward scattering
amplitude \begin{math} g_1  = U(2k_F  + \Delta k + k - k')\end{math},
which can be assumed to be approximately constant for \textit{k, k'
}and \textit{\begin{math} \Delta{}\end{math}k} on the range of the
cutoff \textit{\begin{math} \Lambda{}\end{math}} in most cases. 
Therefore, using (\ref{rhoR})  and (\ref{bL})
\begin{equation}
H_{\rm backward} \approx \frac{1}{N} \sum_{k,k',\Delta k} g_1 \left(
\rho_{\Delta k}^R 
\rho_{-\Delta k}^L 
+\rho_{\Delta k}^L 
\rho_{-\Delta k}^R 
\right)
\label{e79}
\end{equation}
which has the same effect as the dispersion scattering equation (\ref{e77}) above
and therefore can be absorbed right away by a redefinition of the
corresponding scattering amplitude \textit{g$_{2}$}.  For electrons
with spin the backward scattering also introduces a spin-flip term,
which can be taken into account by a renormalization group treatment,
but will not be addressed here.

{~\\ \bf Umklapp scattering}

Finally, there is the possibility of Umklapp scattering of two left
movers into two right movers and vice versa in case that
\begin{math} \Delta k =  \pm \left( {4k_F  + \delta k} \right) =  \pm
\left( {2\pi  + \delta k} \right)\end{math}.  This Umklapp scattering
(denoted by \textit{$g_{3}$}) is obviously only present if the special
condition for half-filling
\textit{$4k_{F}=2 \pi{}$} is fulfilled.  The
resulting operator can then also be bosonized with the help of the
fermion field expressions (\ref{bosR}) and (\ref{bosL}), 
but we will not consider this special
case here.

It should be noted here that the bosonization of the various
scattering terms can also be derived from the interaction Hamiltonian
in real space (\ref{e5}) by using the linearization formula (\ref{psi})  
together with (\ref{rhoLR}) if we
assume a short range potential \begin{math} U(m)\end{math} (as can be
verified as an exercise).  However, the derivation in momentum
space has the advantage that the \begin{math} k\end{math}-dependence of
the scattering amplitudes \textit{g$_{2}$} and\textit{ g$_{4}$ }can be
preserved.\textit{ }

\subsection{The Luttinger Liquid parameter: Boguliobov
transformation }

The bosonized interaction Hamiltonian in (\ref{e77}) or (\ref{e78})  is much easier to treat
than the fermionic version (\ref{e5}), because it is represented by terms that
only involve a product of two boson operators (bilinear) as opposed to
four fermion operators in (\ref{e5})  and (\ref{e6}).  In fact, it is well known how to
solve such problems by a suitable redefinition of the bosonic operators
so that the interaction term has the same form as the original
Hamiltonian, which only contains counting operators.  This procedure is
called a Boguliubov transformation and is given by the general ansatz
\begin{equation} \begin{gathered}
\tilde b_n^R  = b_n^R \cosh \theta  - b_n^{L \dagger} \sinh \theta 
\hfill \\
\tilde b_n^L  = b_n^L \cosh \theta  - b_n^{R \dagger } \sinh \theta 
\hfill \\
\end{gathered}
\label{e80}
\end{equation} 
which defines new operators \begin{math} \tilde b_n^R {\text{ and
}}\tilde b_n^L \end{math}.  It is left as an exercise to show that
those new operators also obey the canonical commutation relations (\ref{e29}).

We notice that the \textit{g$_{4}$} interaction in (\ref{e77}) or (\ref{e78})  is already in
the non-interacting form of equations (\ref{bosH3}) and (\ref{bosH2}) 
and can be absorbed right away by defining a new Fermi
velocity
\begin{equation} \tilde v_F  = v_F  + \frac{{g_4 }}
{\pi }\end{equation} 
Therefore the complete Hamiltonian can be written  as
\begin{equation} \begin{gathered}
H = a\tilde v_F \int_0^\ell  {dx\;\left( {(\partial _x \phi _R
)^2  + (\partial _x \phi _L )^2  + \frac{{2g_2 }}
{{\tilde v_F \pi }}\partial _x \phi _R \partial _x \phi _L }
\right)}  \hfill \\
\quad {\kern 1pt}  = \sum\limits_{n = 1}^\infty  {\frac{{2\pi \tilde
v_F }}
{N}n\left( {b_n^{R \dagger } b_n^R  + b_n^{L \dagger} b_n^L  + \frac{{g_2 }}
{{\tilde v_F \pi }}\left( {b_n^{R \dagger } b_n^{L \dagger}  + b_n^R b_n^L }
\right)} \right)}  \hfill \\
\quad \quad \quad \quad  + \frac{{\tilde v_F }}
{N}\left( {Q_R^2  + Q_L^2  + \frac{{2g_2 }}
{{\tilde v_F \pi }}Q_R^{} Q_L^{} } \right) \hfill \\
\end{gathered} \label{e81} \end{equation} 
(where \textit{g$_{2}$} is assumed to contain also any possible
backward scattering contributions (\ref{e79})).  It is clear now that we want to
use the canonical transformation (\ref{e80})  in order to get rid of the cross
terms \begin{math} b_n^{R \dagger } b_n^{L \dagger}  + b_n^R b_n^L \end{math}. 
This can be done by comparing with the following expression (up to a
constant)
\begin{equation} \tilde b_n^{R \dagger } \tilde b_n^R  + \tilde b_n^{L \dagger} \tilde
b_n^L  = \left( {\cosh ^2 \theta  + \sinh ^2 \theta } \right)\left(
{b_n^{R \dagger } b_n^R  + b_n^{L \dagger} b_n^L } \right) - 2\cosh \theta \sinh
\theta (b_n^{L \dagger} b_n^{R \dagger }  + b_n^L b_n^R )\end{equation} 
which yields the following result for the rotation angle $\theta$
\begin{equation} K = e^{2\theta }  = \sqrt {\frac{{1 - g_2 /v_F \pi }}
{{1 + g_2 /v_F \pi }}}
\label{e82}
\end{equation} 
where \begin{math} K\end{math} is the so called \textit{Luttinger
liquid parameter}.  It is straightforward to verify this result using
the following formulas
\begin{equation} \sinh \theta  = \frac{1}
{2}\left( {\sqrt K  - 1/\sqrt K } \right) \hspace{15pt} 
 \cosh \theta  = \frac{1}
{2}\left( {\sqrt K  + 1/\sqrt K }
\right) \hspace{15pt}\hspace{15pt}\sinh \theta  +
\cosh \theta  = \sqrt K \nonumber \end{equation} 
\begin{equation} \sinh ^2 \theta  + \cosh ^2 \theta  = \left( {K + 1/K}
\right)/2 \hspace{15pt}\hspace{15pt}2\sinh \theta
\cosh \theta  = \left( {K - 1/K} \right)/2
\end{equation} 
Likewise, we can define new number operators
\begin{equation} \begin{gathered}
\tilde Q_R  = Q_R \cosh \theta  - Q_L \sinh \theta  \hfill \\
\tilde Q_L  = Q_L \cosh \theta  - Q_R \sinh \theta  \hfill \\
\end{gathered} \label{e83}
\end{equation} 
after which the Hamiltonian is in the standard form  again
\begin{equation} \begin{gathered}
H = \sum\limits_{n = 1}^\infty  {\frac{{2\pi \tilde v_F }}
{N}n\left( {\tilde b_n^{R \dagger } \tilde b_n^R  + \tilde b_n^{L \dagger} \tilde
b_n^L } \right) + \frac{{\tilde v_F }}
{N}\left( {\tilde Q_R^2  + \tilde Q_L^2 } \right)}  \hfill \\
\quad {\kern 1pt}  = a\tilde v_F \int_0^\ell  {dx\;\left( {(\partial
_x \tilde \phi _R )^2  + (\partial _x \tilde \phi _L )^2 }
\right),}  \hfill \\
\end{gathered} \label{e84}
\end{equation} 
where we have also defined new fields \begin{math} \tilde \phi _R
{\text{ and }}\tilde \phi _L \end{math}  analogous to equations (\ref{e80}) and
(\ref{e83}).  More compactly we can summarize these formulas in a canonical
rescaling equation for the difference and the sum of the left and right
moving fields
\begin{eqnarray}
\tilde \phi_R - \tilde \phi_L  & = &  \sqrt{K} (\phi_L- \phi_R) \nonumber \\
\tilde \phi_R + \tilde \phi_L  & = &  \frac{1}{\sqrt{K}} (\phi_L+ \phi_R) \label{e85}
\end{eqnarray}
This implies also that the creation operators \begin{math} \tilde
\phi _0^R {\text{ and }}\tilde \phi _0^L \end{math}  are
transformed, but it should be remembered that the quantization relation
 must still be fulfilled and particle excitations are performed by the
\textquotedblleft{}old\textquotedblright{} operator \begin{math} \exp
\left( { - i\sqrt {4\pi } \phi _0^R } \right)\end{math} .

In summary, we have therefore solved an interacting problem exactly by
a simple canonical transformation.  Moreover, all interactions can be
described by a single \textit{Luttinger liquid parameter}
$K=e^{2\theta}$ in equation (\ref{e82}).  Even though
\begin{math} K\end{math} can in principle be dependent on the momentum
transfer \textit{\begin{math}k=n2 \pi{}\end{math}/N} it is often
sufficient to just use a constant value for short range interacting
models.  However, it turns out that the expression  is only correct to
lowest order in the actual scattering amplitudes \textit{g$_{2}$} in
typical lattice Hamiltonians like nearest neighbor interactions or the
Hubbard model\cite{16},
 because higher order operators renormalize the interactions. 
Therefore, the actual value of the Luttinger liquid parameter
\begin{math} K\end{math} must almost always be inferred from other
theoretical methods\cite{16,17} or by comparisons with experiments.  For
repulsive interactions we see from equation (\ref{e82})  that
\begin{math} K<1\end{math}.  For \begin{math}K=1 \end{math} we recover
the non-interacting theory.

\subsection{Correlation functions }

To conclude this chapter, we would now like to apply our solution of
the problem in order to calculate 
correlation functions in an interacting model.  Again starting from (\ref{e61})  we
are interested in the correlator of the right moving field
\begin{math} \psi _R  \propto e^{i\sqrt {4\pi } \phi _R (x)}  =
e^{i\sqrt {4\pi } (\tilde \phi _R \cosh \theta  + \tilde \phi _L
\sinh \theta )} \end{math}.  In particular,
\begin{displaymath} 
\begin{gathered}
\left\langle {\psi _R^\dagger  (x)\psi _R (y)} \right\rangle  \propto
e^{2\pi \left[ {\phi _R \left( x \right),\phi _R \left( y
\right)} \right]} \left\langle {e^{ - i\sqrt {4\pi } \left( {\phi _R
\left( x \right) - \phi _R \left( y \right)} \right)} }
\right\rangle  \hfill \\
\quad \quad \quad \quad \quad \quad  = e^{i\pi {\text{sign}}(x -
y)/2} \left\langle {e^{ - i\sqrt {4\pi } (\tilde \phi _R (x) -
\tilde \phi _R (y))\cosh \theta } } \right\rangle \left\langle {e^{
- i\sqrt {4\pi } (\tilde \phi _L (x) - \tilde \phi _L (y))\sinh
\theta } } \right\rangle  \hfill \\
\end{gathered}
\end{displaymath} 
where we have used (\ref{bosR}) and (\ref{e80}).  The transformed operators
\begin{math} \tilde \phi _R {\text{ and }}\tilde \phi _L
\end{math}  have the same canonical mode expansion as before and
therefore the calculation of the correlation functions proceeds as in
section 2.5.

In contrast to equation (\ref{e65}), however, the overall proportionality
constant is now cut-off dependent and not normalized to unity.  The
left moving correlator differs by a minus sign.
\begin{eqnarray} \left\langle {\psi _R^\dagger  (x)\psi _R (y)} \right\rangle 
& \propto &  i{\text{ sign}}(x-y)\left| {2\ell \sin \frac{{\pi}}
{\ell }(x-y)} \right|^{-\cosh^2\theta} 
\left| {2\ell \sin \frac{{\pi}}
{\ell }(x-y)} \right|^{-\sinh^2\theta} \nonumber \\
& \propto &  i{\text{ sign}}(x-y)\left| {2\ell \sin \frac{{\pi}}
{\ell }(x-y)} \right|^{-(K+1/K)/2}
\end{eqnarray} 

As in section 2.5 we can also use the time dependent mode expansion 
in order to calculate the Green's function.  Accordingly, we find
\begin{equation} \left\langle {\psi _R^\dagger  (t)\psi _R (0)} \right\rangle 
\propto i{\text{ sign}}(t)\left| {2\ell \sin \frac{{\pi v_F at}}
{\ell }} \right|^{ - (K + 1/K)/2} \label{e87}
\end{equation} 

A famous result for the density of states can now be obtained in the
thermodynamic limit
\textit{\begin{math} \ell\end{math}\begin{math}\rightarrow{}\end{math}\begin{math}\infty{}\end{math},
}since according to equations (\ref{G})  and (\ref{e87})
\begin{equation} G^R (t,x)\, = \, - i\,\left\langle {\{ \psi (x,t),\psi ^\dagger
 (x,0)\} } \right\rangle \theta (t)\;\mathop  \propto \limits^{\ell 
\to \infty } \;\;t^{ - (K + 1/K)/2}
\end{equation} 
and therefore from  the density of state becomes,
\begin{equation} \rho (\omega ) \propto \int {dt} \,e^{i\omega t} t^{ - (K
+ 1/K)/2}  \propto \omega ^{(K + 1/K)/2 - 1}. 
\label{e88}
\end{equation} 
where we have simply used a rescaling of the integration variable
\textit{x\begin{math} =\omega{}\end{math}t}.  This is the famous result
of the depletion at low frequencies of the single particle spectral
weight in a Luttinger liquid with a characteristic power-law
\begin{equation} \rho (\omega ) \propto \omega ^\alpha  ,\quad \quad \quad
\quad \quad \alpha  = \frac{{K + 1/K - 2}}
{2} > 0
\label{e89}
\end{equation} 

This depletion is one of the main hallmarks which is used in order to
detect Luttinger liquid behavior experimentally with tunneling and
photoemission experiments\cite{7,8,9} as described in section 1.2.  It
must be emphasized again that the proportionality constant is not
known, but tunneling experiments cannot detect the overall amplitude of
the density of states either.  Theoretically it is possible to describe
the proportionality constant of correlators phenomenologically with a
cut-off energy scale \cite{1,2,3,4,5}
or by using a restricted momentum range for the
interactions\cite{18},
but these are just mathematical tools that do not determine the
actual value.  In some microscopic models it is possible to fix the
proportionality constant and the Luttinger liquid parameter by
comparison with other exact methods\cite{16,17,19}.

As mentioned above, the correlators for electrons with spin factorize
in a spin part and a charge part.  In this case the interactions only
transform the charge bosons unless spin-dependent scattering is
considered.  Therefore, we can immediately generalize equation (\ref{e76})  for the
interacting case
\begin{equation} \left\langle {\psi _R^{ \uparrow  \dagger } (x)\psi _R^ \uparrow
 (y)} \right\rangle  \propto i{\text{ sign}}(x - y)\left| {2\ell \sin
\frac{{\pi (x + y)}}
{\ell }} \right|^{ - 1/2} \left| {2\ell \sin \frac{{\pi (x - y)}}
{\ell }} \right|^{ - (K + 1/K)/4} \label{e90} \end{equation} 
and analogously for the time correlations.  The corresponding result
for the power-law depletion of the density of states becomes analogous
to (\ref{e88}) and (\ref{e89})
\begin{equation} \rho (\omega ) \propto \omega ^\alpha  ,\quad \quad \quad
\quad \quad \alpha  = \frac{{K + 1/K - 2}}
{4} > 0\end{equation} 
If a general correlation in space and time is calculated, the two
factors in (\ref{e90}) contain different velocities \begin{math} v_c  > v_s
\end{math}, corresponding to the separate excitations of spin and
charge.

This concludes the elementary introduction to bosonization.  Using the
tools we have developed here the reader is encouraged to explore also
some of the more advanced topics, which can for example be found in the
suggested review articles\cite{1,2,3,4,5}
.

\section{Appendix:  The boson cummulant formula}

The cummulant formula for bosons states that the finite temperature
expectation value of an exponential of a linear combination of boson
creation and annihilation operators can be expressed as the exponential
of an expectation value in the following way
\begin{equation} \left\langle {e^{ \alpha b + \beta b^\dagger  }}
\right\rangle  = e^{\left\langle {(\alpha b + \beta b^\dagger  )^2
} \right\rangle /2}
\label{a1} \end{equation} 
If more than one species of bosons is present, this expression
factorizes on both sides so that it can be used for any linear
combination of bosons \begin{math} f = \sum\limits_n {\left( {\alpha _n
b_n  + \beta _n b_n^\dagger  } \right)} \end{math}.  It is not valid for the
zero-mode operators, however (see Ref.~\cite{15}
for a discussion on finite temperature expectation values of zero
modes).

The right hand side of equation (\ref{a1}) can be evaluated rather easily by use of the
Bose-Einstein distribution for finite temperatures
\textit{\begin{math} \beta{}\end{math}=}1\textit{/k$_{B}$T }in terms of
the corresponding energy quantum
$\hbar \omega$
\begin{displaymath} 
\left\langle {b^\dagger  b} \right\rangle  = \frac{{\sum\limits_{n =
0}^\infty  {e^{ - n\beta \hbar \omega } \left\langle {n\left| {b^\dagger  b}
\right|n} \right\rangle } }}
{{\sum\limits_{n = 0}^\infty  {e^{ - n\beta \hbar \omega } } }} =
\frac{{e^{ - \beta \hbar \omega } }}
{{e^{ - \beta \hbar \omega }  - 1}}
\end{displaymath} 
Therefore,
\begin{equation} \begin{gathered}
\ \ e^{ {\left\langle {(\alpha b + \beta b^\dagger  )^2 }
\right\rangle /2} } = e^{ {\alpha \beta
\left\langle {bb^\dagger   + b^\dagger  b} \right\rangle /2}} 
\hfill \\
 \quad \quad \quad \quad \quad \quad \;{\kern 1pt} 
= e^{\alpha \beta \left\langle {b^\dagger  b} \right\rangle
{\text{ + }}\alpha \beta \left[ {b,b^\dagger  } \right]{\text{/2 }}} 
\hfill \\
 \quad \quad \quad \quad \quad \quad \;{\kern 1pt} 
= e^{\frac{{\alpha \beta q}}
{{1 - q}}{\text{ + }}\frac{{\alpha \beta }}
{{\text{2}}}{\text{ }}}  \hfill \\
 \quad \quad \quad \quad \quad \quad \;{\kern 1pt} 
= e^{ - \alpha \beta /2} e^{\frac{{\alpha \beta }}
{{1 - q}}}  \hfill \\
\end{gathered} \label{e92} \end{equation} 
where we have introduced the Boltzmann weight \textit{q =
}exp(\textit{\textendash{}\begin{math} \beta{}\end{math}\begin{math}\hbar{}\end{math}\begin{math}\omega{}\end{math}})\textit{.
}

For the left hand side of (\ref{a1})
we use the Baker Hausdorf formula (\ref{baker}) and
expand the exponential
\begin{equation} \begin{gathered}
\quad \left\langle {e^{\alpha b + \beta b^\dagger  }} \right\rangle  =
\left\langle {e^{\alpha b}e^{\beta b^\dagger  }} \right\rangle 
e^{\alpha \beta \left[ {b^\dagger  ,b} \right]/2}  \hfill \\
 \quad \quad \quad \quad \quad \;\;\, = e^{ - \alpha \beta /2}
\sum\limits_{n,n' = 0}^\infty  {\frac{{\alpha^n \beta^{n'} }}
{{n!n'!}}\left\langle {b^n b^{n'\dagger}  } \right\rangle }  \hfill \\
\quad \quad \quad \quad \quad \;\;\, = e^{ - \alpha \beta /2}
\sum\limits_{n = 0}^\infty  {\frac{{(\alpha \beta )^n }}
{{(n!)^2 }}\frac{{\sum\limits_{m = 0}^\infty  {q^m \left\langle
{m\left| {b^n b^{n\dagger} } \right|m} \right\rangle } }}
{{\sum\limits_{m = 0}^\infty  {q^m } }}}  \hfill \\
 \quad \quad \quad \quad \quad \;\;\, = (1 - q)e^{ - \alpha
\beta /2} \sum\limits_{n = 0}^\infty  {\frac{{(\alpha \beta )^n }}
{{(n!)^2 }}\sum\limits_{m = 0}^\infty  {q^m \left\langle {m\left| {b^n
b^{n\dagger} } \right|m} \right\rangle } }  \hfill \\
\end{gathered} \label{e93} \end{equation} 
where we have used the standard expression for a temperature
expectation value with the Boltzmann weight \begin{math} q\end{math}. 
For a harmonic oscillator it is well known that the expectation value
in the \begin{math} m\end{math}$^{th}$ excited state is given by
\begin{displaymath} 
\left\langle {m\left| {b^n b^{n\dagger} } \right|m} \right\rangle  =
\frac{{(m + n)!}}
{{m!}}
\end{displaymath} 
(e.g.~by repeated use of \begin{math} b^\dagger  \left| m \right\rangle  =
\sqrt {m + 1} \left| {m + 1} \right\rangle \end{math} ).  Finally, we
insert into equation (\ref{e93})  the following Taylor expansion around
\begin{math} q=0\end{math}
\begin{displaymath} 
(1 - q)^{ - n - 1}  = \sum\limits_{m = 0}^\infty  {q^m } \frac{{(m +
n)!}}
{{n!m!}}
\end{displaymath} 
(which can be verified by repeated differentiation).  Therefore,
equation  (\ref{e93}) becomes
\begin{displaymath} 
\begin{gathered}
\left\langle {e^{\alpha b + \beta b^\dagger } } \right\rangle  = (1 -
q)e^{ - \alpha \beta /2} \sum\limits_{n = 0}^\infty  {\frac{{(\alpha
\beta )^n }}
{{n!}}(1 - q)^{ - n - 1} }  \hfill \\
 \quad \quad \quad \quad \quad \;\;\, = e^{ - \alpha \beta /2}
\sum\limits_{n = 0}^\infty  {\frac{1}
{{n!}}\left( {\frac{{\alpha \beta }}
{{1 - q}}} \right)^n }  \hfill \\
 \quad \quad \quad \quad \quad \;\;\, = e^{ - \alpha \beta /2}
\exp \left( {\frac{{\alpha \beta }}
{{1 - q}}} \right) \hfill \\
\end{gathered}
\end{displaymath} 
which is exactly the same expression as on the left hand side (\ref{e92})  and
therefore concludes the proof of the cummulant theorem.  Note that at
\begin{math} T=0\end{math}, we have\textit{ q =
}exp(\textit{\textendash{}\begin{math} \beta{}\end{math}\begin{math}\hbar{}\end{math}\begin{math}\omega{}\end{math}})\textit{\begin{math}\rightarrow{}\end{math}}0,\textit{
}so that
\begin{displaymath} 
\left\langle {e^{\alpha b + \beta b^\dagger  }} \right\rangle  = 
e^{\left\langle {(\alpha b + \beta b^\dagger  )^2 } \right\rangle
/2 } = \exp \left( {\frac{{\alpha \beta }}
{2}} \right)
\end{displaymath} 
as already shown in equation (\ref{e63}) .

\end{document}